\documentclass[aps,pra,twocolumn]{revtex4}  

\usepackage{epsfig}
\newcommand{\be}{\begin{eqnarray}}  
\newcommand{\ee}{\end{eqnarray}}  
  
\def\la{\langle}  
\def\ra{\rangle}

\def\ua{\uparrow}
\def\da{\downarrow}

\newcommand{\mat}{\left( \begin{array}{cc}}
\newcommand{\rix}{\end{array} \right)}

\begin{document}  
  
\title{Quantum random walks - an introductory overview}  
  
\author{J. Kempe}  
\affiliation {CNRS-LRI, UMR 8623, Universit\'e de Paris-Sud, 91405 Orsay,  
France \\ Computer Science Division and Department of Chemistry, University of California, Berkeley }  

\begin{abstract}  
This article aims to provide an introductory survey on quantum random walks.  Starting from a physical effect to illustrate the main ideas we will introduce quantum random walks, review some of their properties and outline their striking differences to classical walks. We will touch upon both physical effects and computer science applications, introducing some of the main concepts and language of present day quantum information science in this context. We will mention recent developments in this new area and outline some open questions.
\end{abstract}

\date{\today}  
\maketitle

\vskip.5cm
%\begin{flushright}
\hskip4cm \begin{minipage}{4cm}{\em God does not play dice.}  \\
\indent \hskip1cm  {\em Albert Einstein}
\end{minipage}
%\end{flushright}

\section{Overview}

Ever since the discovery of quantum mechanics people have been puzzled by the counter-intuitive character of the laws of nature. Over time we have learned to accept more and more effects that are unimaginable in a classical Newtonian world. Modern technology exploits quantum effects both to our benefit and detriment - among the memorable examples we should cite laser technology and not omit the atomic bomb. 

In recent years interest in quantum information theory has been generated by the prospect of employing its laws to design devices of surprising power \cite{Nielsen:book}. New ideas include quantum cryptography \cite{Hughes:95a,Gisin:02a} and quantum computation. In 1994 Peter Shor \cite{Shor:94a} discovered a quantum algorithm to factor numbers efficiently (that is in time that grows only polynomial with the length of the number to be factored), which has unleashed a wave of activity across a broad range of disciplines: physics, computer science, mathematics and engineering. 

This fruitful axis of research has uncovered many new effects that are strikingly different from their classical counterparts, both from the physical point of view as well as from a computer science and communication theory perspective. Over time these communities have gained a greater understanding of the  concepts and notions of the other. The idea that information cannot be separated from the physical devise that is carrying it ('Information is physical') has firmly settled among them and has led to fascinating new insights. Acquaintance with the basic notions of each of these fields seems instrumental in the understanding of modern quantum information processing.

In this article we will follow the trajectory of one of the many surprising aspects of quantum information; it is dedicated to quantum random walks. We will give a thorough introduction to the necessary terminology without overburdening the reader with unnecessary mathematics. Starting with a very intuitive and physical example we will step by step introduce the language and notation of present day quantum information science. We will present the necessary background from computer science needed for a physicist to understand and appreciate the developments and results, assuming some rudimentary background of quantum mechanics, but no knowledge of computer science or quantum information theory. An excellent comprehensive introduction to quantum information and computation can be found in \cite{Nielsen:book}.

In this journey at the interface of several traditional disciplines we will follow quantum random walks both through physics and computer science. This tour through quantum information science will allow us to survey some of its most pertinent concepts and ideas today, like quantum algorithms, quantum computing machines, speed-ups, physical implementation, quantum circuits and decoherence. It will be our mission to show how physical phenomena translate into new computer science algorithms and vice versa.
The idea is that a reader familiar with the standard idiom of quantum mechanics but unaccustomed to the language of 'qubits' and 'gates' will learn about quantum information science while learning about quantum random walks and their fascinating behavior. We will survey some recent developments and results omitting most proofs but trying to develop an intuition. This article aims to bring the interested reader to a point where he or she can read and understand current research articles on the topic and develop an understanding for the interesting problems and open questions in the area.

The structure of this article is as follows: First we give some physical intuition for random walks, providing a general flavor of the phenomenon in a physical setting (Section \ref{Sec:gentle}). This is followed by a more rigorous definition together with some necessary terminology to introduce the two main models of quantum random walks in Section \ref{Sec:walk}. We then present some computer science and probability background and mention some of the important algorithmic results stemming from quantum random walks in Section \ref{Sec:cs}. Section \ref{Sec:implementation} switches back to physics and studies how these random walks could be implemented in some real physical system. Finally we deal with the more philosophical question of how the classical world emerges from quantum behavior via decoherence  using random walks as our example (Section \ref{Sec:decoherence}). All along the way we will outline open questions and future directions.

\section{A gentle introduction}\label{Sec:gentle}

To illustrate quantum random walks and give an intuition for what follows we start with an example. This example - which might be thought of as a precursor to later models - is taken from the work of three physicists in 1993,  Y. Aharonov, L. Davidovich and N. Zagury \cite{Aharonov:93a}. Their work  for the first time coins the term 'quantum random walk'. 
%It will allow us to conveniently introduce the language and notation of present-day quantum information theory.

Imagine a particle  on a line whose position is described by a wave-packet $|\psi_{x_0} \ra$ localized around a position $x_0$, i.e. the function $\la x|\psi_{x_0}\ra$ corresponds to a wave-packet centered around $x_0$. Let $P$  be the momentum operator. The translation of the particle, corresponding to a step of length $l$ can be represented by the unitary operator $U_l=exp(-iPl/\hbar)$, so that $U_l|\psi_{x_0}\ra=|\psi_{x_0-l}\ra$. Now let us also assume that the particle has a spin-$\frac{1}{2}$ degree of freedom. Let $S_z$ represent the operator corresponding to the $z$ component of the spin and denote the eigenstates of $S_z$ by $|\ua\ra$ and $|\da\ra$, so that $S_z |\ua\ra=\frac{\hbar}{2} |\ua\ra$ and $S_z| \da\ra=- \frac{\hbar}{2} |\da \ra$. From now on we will set $\hbar=1$ to simplify the notation. 

A spin-$\frac{1}{2}$ particle is usually described by a $2$-vector $|\Psi\ra=(|\tilde{\psi}^{\ua}\ra,|\tilde{\psi}^{\da}\ra)^T$, where the first part is the component of the wave-function of the particle in the spin- $|\ua\ra$ space and the second one is the component in the $|\da\ra$-space. 
Normalization requires that $\||\tilde{\psi}^\ua\ra \|^2+\||\tilde{\psi}^\da\ra\|^2=1$.
 To emphasize the tensor-structure of the space of the particle we will write this in a slightly different but equivalent way as $|\Psi \ra = \alpha^\ua |\ua\ra \otimes |\psi^\ua \ra + \alpha^\da |\da\ra \otimes |\psi^\da\ra$, where we normalize the two wave-functions $\la \psi^\ua|\psi^\ua\ra=\la \psi^\da|\psi^\da\ra=1$, so that $|\alpha^\ua|^2+|\alpha^\da|^2=1$. The tensor product '$\otimes$' separates the two degrees of freedom, spin and space, and will allow us to view the resulting correlations between these two degrees of freedom more clearly. The time development corresponding to a translation by $l$ on the larger state-space of the spin-$\frac{1}{2}$ particle can now be described by the unitary operator $U=exp(-2iS_z\otimes Pl)$. This operator induces a kind of conditional translation of the particle depending on its internal spin-degree of freedom. In particular if the spin of the particle is initially in the state $|\ua\ra$, so that its wave-function is of the form $|\ua\ra \otimes |\psi^{\ua}_{x_0}\ra$, then application of $U$ transforms it to   $|\ua\ra \otimes |\psi^{\ua}_{x_0-l}\ra$ and the particle will be shifted to the right by $l$. If the spin of the particle is in the state $|\da\ra$, i.e. the total wave-function is given by $|\da\ra \otimes |\psi^{\da}_{x_0}\ra$, then the translation operator will transform it to $|\da\ra \otimes |\psi^{\da}_{x_0+l}\ra$ and the particle will be shifted to the left. More interesting behavior occurs when the initial spin state of the particle, localized in $x_0$, is not in an eigenstate of $S_z$, but rather in a superposition 
\be \label{Eq:init}
|\Psi_{in}\ra= (\alpha^\ua |\ua\ra + \alpha^\da |\da\ra) \otimes |\psi_{x_0}\ra. 
\ee
Application of the translation operator $U$ will induce a superposition of positions
\be
U|\Psi_{in}\ra = \alpha^\ua |\ua\ra \otimes |\psi_{x_0-l}\ra + \alpha^\da |\da\ra \otimes |\psi_{x_0+l}\ra.
\ee
If at this point we decide to {\em measure} the spin in the $S_z$ basis, the particle will be either in the state $|\ua\ra \otimes |\psi_{x_0-l}\ra$, localized around $x_0+l$ with probability $p^\ua=|\alpha^\ua|^2$ or in the state $|\da\ra \otimes |\psi_{x_0-l}\ra$, localized around $x_0-l$ with probability $p^\da=|\alpha^\da|^2$. This procedure corresponds to a (biased) random walk of a particle on the line: we can imagine that a biased coin with probabilities of head/tail $p^\ua /p^{\da}$ is flipped. Upon head the particle moves right and upon tail the particle moves left. After this step the particle is on average displaced by $l(p^\ua-p^\da)$. If we repeat this procedure $T$ times (each time measuring the spin in the basis $\{|\ua\ra,|\da\ra\}$ and re-initializing the spin in the state $\alpha^\ua |\ua\ra + \alpha^\da |\da\ra$), the particle will be displaced on average by an amount $\la x \ra=Tl(p^\ua-p^\da)=Tl(|\alpha^\ua|^2-|\alpha^\da|^2)$ and the variance of its distribution on the line will be $\sigma^2=4Tl|\alpha^\ua|^2 |\alpha^\da|^2=4Tl^2p^\ua p^\da$. This is exactly what we obtain if the particle performs a (biased) random walk on the line. We can imagine that the spin of the particle takes the role of a coin. This coin is 'flipped' and the outcome determines the direction of the step in the random walk of the particle.

Now let us consider the following modification of this procedure. Instead of measuring the spin in the eigenbasis of $S_z$ we will measure it in some rotated basis, given by two orthogonal vectors $\{|s_+\ra,|s_-\ra\}$. Alternatively we can {\em rotate} the spin by some angle $\theta$ before measuring it in the $S_z$ eigenbasis. Let us be more formal and set up some more of the language used in quantum information theory. If we identify 
\be 
|\ua\ra=\left( \begin{array}{c} 1 \\ 0 \end{array} \right) \quad |\da\ra =\left( \begin{array}{c} 0 \\ 1 \end{array} \right)
\ee 
we can write
\be \label{Eq:Sz}
S_z=\frac{1}{2} \left( \begin{array}{cc} 1 & 0 \\ 0 & -1 \end{array} \right) = \frac{1}{2} (|\ua\ra\la \ua|-|\da\ra \la \da|).
\ee
A rotation of the spin can be described by the matrix\footnote{Note that this is not the most general unitary transformation on a two-level system. A general (normalized) unitary is given by two parameters $(\theta,\phi)$.}
\be
R(\theta)=\left( \begin{array}{cc} \cos \theta & -\sin \theta \\ \sin \theta & \cos \theta \end{array} \right).
\ee
Let us denote the measurement in the $S_z$ basis by $M_z$.  To explore the effect of the suite of operations $M_zR(\theta)U$ on the initial state $|\Psi_{in}\ra$ we have to slightly rewrite the operator $U$:
\begin{eqnarray} \label{Eq:shift}
U &=& e^{-2iS_z\otimes Pl}=e^{-i (|\ua\ra\la \ua|-|\da\ra \la \da|)\otimes Pl}\nonumber \\ &=&(|\ua\ra\la \ua|\otimes e^{-i Pl})(|\da\ra\la \da|\otimes e^{i Pl}).
\end{eqnarray}
The second equality follows from the fact that $|\ua\ra \la \ua| \otimes Pl$ and $|\da\ra \la \da| \otimes Pl$ commute\footnote{The equality $\exp(A+B)=\exp(A)\exp(B)$ is true if $A$ and $B$ commute, i.e. $AB=BA$.} together with the observation that $|\ua\ra \la \ua|$ and $|\da\ra \la \da|$ are projectors, so $|\ua\ra \la \ua|^k=|\ua\ra \la \ua|$ and $|\da\ra \la \da|^k=|\da\ra \la \da|$ when we expand the exponential. Now 
\begin{equation}
U |\Psi_{in}\ra= (\alpha^\ua |\ua\ra \otimes e^{-i Pl}+\alpha^\da|\da\ra \otimes e^{i Pl})|\psi_{x_0}\ra 
\end{equation}
and applying $R(\theta)$ from the left gives
\begin{equation}
\begin{array}{l}
\left[ (\alpha^\ua \cos \theta e^{-iPl}-\alpha^\da \sin \theta e^{i Pl})|\ua\ra \right.  \\ \left. +(\alpha^\ua \sin \theta e^{-iPl}+\alpha^\da \cos \theta e^{i Pl})|\da\ra \right] \otimes |\psi_{x_0}\ra. 
\end{array}
\end{equation}
If the width $\Delta x$ of the initial wave packet is much larger than the step length $l$ we can approximate $\exp(\pm iPl) |\psi_{x_0}\ra \approx (I\pm iPl)|\psi_{x_0}\ra$. With this in mind we can establish the state of the particle after the measurement $M_z$:
\begin{equation}
M_zR(\theta)U|\Psi_{in}\ra = \left\{ \begin{array}{c} |\ua\ra \otimes (I-iPl \delta^\ua )|\psi_{x_0}\ra    \\ |\da\ra \otimes (I-iPl \delta^\da)|\psi_{x_0}\ra  \end{array} \right.
\end{equation}
with probabilities
\be
p^\ua=|\alpha^\ua \cos \theta - \alpha^\da \sin \theta |^2  \quad p^\da=|\alpha^\ua \sin \theta + \alpha^\da \cos \theta |^2
\ee
and displacements 
\be
l \delta^\ua:=l\frac{\alpha^\ua \cos \theta + \alpha^\da \sin \theta}{\alpha^\ua \cos \theta - \alpha^\da \sin \theta} \quad l\delta^\da :=l\frac{\alpha^\ua \sin \theta - \alpha^\da \cos \theta}{\alpha^\ua \sin \theta + \alpha^\da \cos \theta}.
\ee
If the width $\Delta x$ of the initial wave packet is much larger than $l \delta^\ua$ and $l \delta^\da$, we can again approximate $(I-iPl\delta^{\ua,\da})|\psi_{x_0}\ra \approx \exp(-iPl \delta^{\ua,\da})|\psi_{x_0}\ra= |\psi_{x_0-l \delta^{\ua,\da}}\ra$. 

Now it is crucial to note that the displacement of the particle in one of the two cases, $l \delta^\ua$ say, can be made much {\em larger} than $l$. For instance we may choose $\tan \theta = |\alpha^\ua / \alpha^\da|(1+\epsilon)$ with $l/ \Delta x \ll |\epsilon| \ll 1$. Then the displacement of the particle in case we measured $|\ua\ra$ will be $l \delta^\ua \approx -2l / \epsilon$ which can be several orders of magnitude larger than $l$. However this event is 'rare': the probability to measure $|\ua\ra$ is $p^\ua \approx |\alpha^\ua \alpha^\da|^2 \epsilon ^2$. For the same choice of parameters the probability to measure $|\da\ra$ is $p^\da \approx 1- |\alpha^\ua \alpha^\da|^2 \epsilon ^2$ and the displacement in that event is $l \delta^\da \approx l (|\alpha^\ua|^2 - |\alpha^\da|^2) +O(l \epsilon)$. For the average displacement  $p^\ua l \delta^\ua + p^\da l \delta^\da$ we get as before $l (|\alpha^\ua|^2 - |\alpha^\da|^2)$, and similarly the variance of the walk will be unchanged. We have the effect that even though our translation operation displaces only by $l$, in some (rare) cases the particle will jump much further than $l$. The effect is illustrated in Fig. \ref{Fig:effect} below.

%\begin{center}
%\vbox{
\begin{figure}[t]
\epsfxsize=8cm
\epsfbox{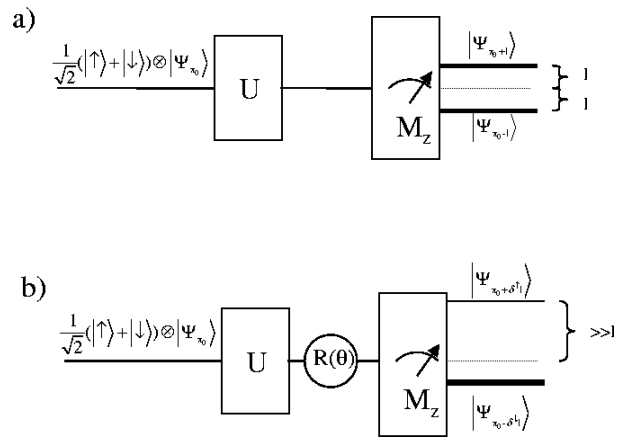}
\caption{a) The particle is localized in position $x_0$ and enters the conditional displacement box $U$. Here depending on its spin degree of freedom it is shifted up or down by $l$. Upon exit its spin is measured in the $z$ basis ($M_Z$). With probability $1/2$ it is found displaced by $l$ resp. $-l$. b) The same set-up except that a rotation is applied to the spin after the particle exits the $U$-box. After the measurement $M_z$ of the spin the particle now can be shifted up by far more than $l$ (with small probability). With large probability it is shifted by less than $l$ in the other direction.}
\label{Fig:effect}
\end{figure}
%}
%\end{center} 

This is strikingly different from any classical behavior. Quantum mechanics allows us to post-select events that cannot be observed classically. Even though these events are rare events we cannot create a classical set-up that allows us to observe such an effect. This 'Gedankenexperiment' unveils some of the 'weirdness' of quantum mechanics. In what follows we will expand  and show how variations on this phenomenon might be useful in the context of modern quantum information processing.

\section{The quantum random walk} \label{Sec:walk} 

Since the work by Aharonov et al. \cite{Aharonov:93a} mentioned in the previous section two models of quantum random walks have been suggested. They can be viewed as a generalization of the effect we have described above in one way or other, yet they differ in several aspects both from the example in Section \ref{Sec:gentle} as well as from each other. In this section we will define both models and describe some results on the difference in behavior of classical and quantum walk.

The key idea behind quantum random walks is to iterate the walk we have introduced in the previous section {\em without} resorting to intermediate measurements.  In other words we will repeat the succession of unitary translation $U$ and rotation $R(\theta)$ without the measurement $M_z$ at intermediate time steps. Furthermore we will discretize the position space of the particle, or in other words we put the particle on a lattice or a graph. Usually this space will be finite. A discrete and finite state space is important in the context of simulability on a (finite) computer. A quantum computer works with discrete registers, i.e. its state-space is a Hilbert space of large but finite dimension. Discretizing the quantum walk will allow us to map it to a computation by such a machine.

Apart from the interesting new physics that we will uncover in the study of quantum random walks we will want to employ some of the 'weird' effects of the walk provided by quantum mechanics to enhance the computing power at our disposal. The idea is that a quantum computer can implement (simulate) the quantum random walk efficiently and use this simulation to solve certain computational tasks. In this vein we hope to use properties of quantum random walks to find more efficient algorithms on a quantum computer. 

\subsection{The discrete model}\label{Sec:discrete}

Let us define the formal set-up for a first model, the so called discrete time quantum random walk. Interestingly this model appears already in the works of Feynman \cite{Feynman:65a} with the discretization of the Dirac equation in mind. In the era of quantum information it was rediscovered in works by Meyer \cite{Meyer:96a,Meyer:96b} in connection with quantum cellular automata and by Watrous \cite{Watrous:unp} in the context of halting of quantum Turing machines and in a slightly modified version (with measurements) in \cite{Watrous:01a} thinking about space-bounded quantum computation. As a possible computational tool and in a formal way it has been introduced and analysed in 2001 by Aharonov, Ambainis, Kempe and Vazirani \cite{Aharonov:01a} (on general graphs) and by Ambainis et al. \cite{Ambainis:01b} (on the line).

We will define our model in one dimension, on the line or the circle. However the definitions carry over to the general case with slight modifications, which we will mention later.

Let ${\cal H}_P$ be the Hilbert space spanned by the positions of the particle. For a line with grid-length $1$ this space is spanned by basis states $\{|i\ra: i \in {\bf Z} \}$; if we work on a circle of size $N$ we have ${\cal H}_P=\{|i\ra: i=0 \ldots N-1\}$. States in this basis will take the role of the wave function $|\psi\ra$ of Section \ref{Sec:gentle} with  $|i\ra$ corresponding to a particle localized in position $i$. We will not be concerned any more about the width of the distribution of the wave function - our model describes unitary transformations of states of a finite Hilbert space ${\cal H}$ and the notion of a particle is only used to guide our intuition from now on. Yet later, when we talk about the physical implementation of the quantum random walk, we will see that we can approximate the states of this mathematical model by particles (atoms, photons, etc.) with wave functions of finite width. 

The position Hilbert space ${\cal H}_P$ is augmented by a 'coin'-space ${\cal H}_C$ spanned by two basis states $\{|\ua\ra,|\da \ra\}$, which take the role of the spin-$\frac{1}{2}$ space in the previous section. States of the total system are in the space ${\cal H}={\cal H}_C \otimes {\cal H}_P$ as before. The conditional translation of the system (taking the role of $\exp(-2iS_z \otimes Pl)$) can be described by the following unitary operation
\be
S=|\ua\ra \la \ua | \otimes \sum_i |i+1\ra \la i|+|\da\ra \la \da | \otimes \sum_i |i-1\ra \la i|
\ee
where the index $i$ runs over $Z$ in the case of a line or $0 \leq i \leq N-1$ in the case of a circle. In the latter case we always identify $0$ and $N$, so all arithmetic is performed modulo $N$. $S$ transforms the basis state $|\ua \ra \otimes |i\ra$ to $|\ua\ra\otimes |i+1\ra$ and $|\da \ra \otimes |i\ra$ to $|\da \ra \otimes |i-1\ra$. 

\begin{figure}[t]
\epsfxsize=8cm
\epsfbox{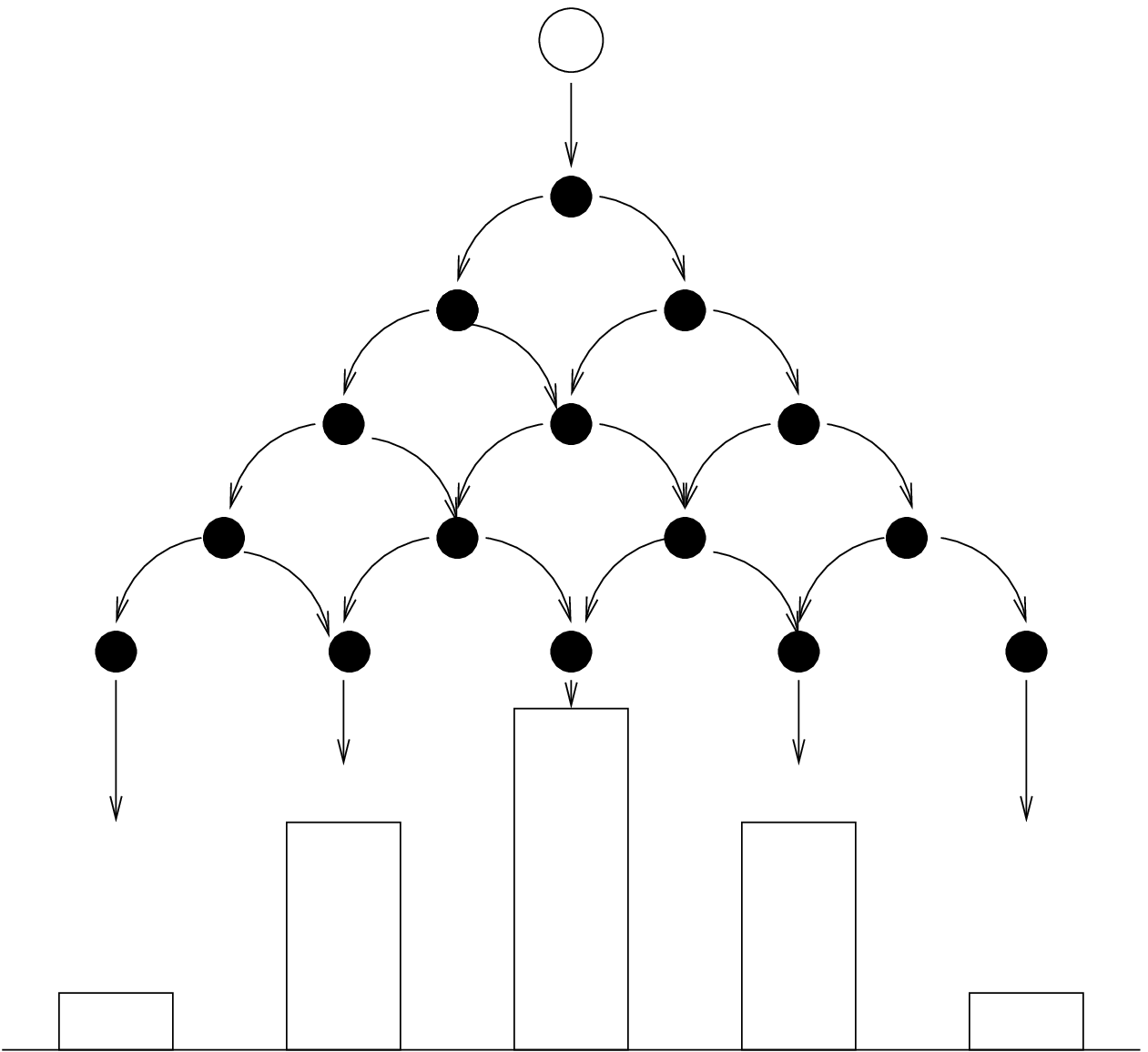}
\caption{Galton's Board (Quincunx): \\ The Quincunx is a device which allows a bead to drop through an array of pins stuck in a board. The pins are equally spaced in a number of rows and when the bead hits a pin it is equally likely to fall to the left or the right. It then lands on a pin in the next row where the process is repeated. After passing through all rows it is collected in a slot at the bottom. The distribution of beads approaches a Gaussian after many rows of pins.}
\label{Fig:1}
\end{figure}

The first step of the random walk is a rotation in the coin-space, which in analogy to the classical random walk we will call 'coin-flip' $C$. The unitary transformation $C$ is very arbitrary and we can define a rich family of walks with different behavior by modifying $C$. For the moment we will want $C$ to be 'unbiased' in the following sense. Assume we would initialize the walk in a localized state $|0\ra$ with the coin in one of the basis states, say $|\ua\ra$. If we chose to measure the coin-register of the walk in the (standard) basis $\{|\ua\ra,|\da \ra \}$ after one iteration (coin-flip $C$ followed by translation $S$) we want to obtain the classical probability distribution of the unbiased walk, namely a translation to the right ($|1\ra$) with probability $\frac{1}{2}$ and a step to the left ($|-1\ra$) otherwise. A frequently used balanced unitary coin is the so called Hadamard coin $H$
\be
H=\frac{1}{\sqrt{2}} \left( \begin{array}{cc} 1 & 1 \\ 1 & -1 \end{array} \right).
\ee
It is easy to see that the Hadamard coin is balanced:
\begin{eqnarray}\label{Eq:rightwalk}
|\ua\ra \otimes |0\ra &\stackrel{H}{\longrightarrow}&\frac{1}{\sqrt{2}}(|0\ra + |1 \ra) \otimes |0\ra \nonumber \\ &\stackrel{S}{\longrightarrow}& \frac{1}{\sqrt{2}}(|\ua\ra \otimes |1\ra + |\da \ra \otimes |-1\ra).
\end{eqnarray}
Measuring the coin state in the standard basis gives each of $\{|\ua\ra \otimes |1\ra,|\da \ra \otimes |-1\ra\}$ with probability $1/2$. After this measurement there is no correlation between the positions left. If we continued the quantum walk with such a measurement at each iteration we obtain the plain classical random walk on the line (or on the circle). The distribution of this walk can be well illustrated with the help of Galton's board (also called Quincunx) in Fig. \ref{Fig:1}. Its limiting distribution on the line (for large number of iterations $T$) approaches a Gaussian distribution with mean zero and variance $\sigma^2=T$. On the circle this distribution approaches the uniform distribution (more on convergence of random walks on finite graphs in Sec. \ref{Sec:cs}).

%\begin{center}
%\vbox{
%}
%\end{center}
In the quantum random walk we will of course not measure the coin register during intermediate iterations, but rather keep the quantum correlations between different positions and let them interfere in subsequent steps. This interference will cause a radically different behavior of the quantum walk\footnote{Very similar ideas - to iterate the random walk and study its interference pattern -  have already been suggested in 1997 independently by Harmin \cite{Harmin:97a} (in a model for time evolution of a system with many potential-energy curves that cross one another) and in 1999 by Bouwmeester et al. \cite{Bouwmeester:99a} (in an optical setting which deals with spectral diffusion of a light wave). If we extract the relevant details in both cases, rather similar effects to the behavior of our quantum walk occur. Yet the scope and focus of these works differs from what is described here. In the context of cellular automata Meyer \cite{Meyer:96a,Meyer:96b} observes similar interference patterns. }. In particular we will see that the limiting distribution of the walk on the line will not approach a Gaussian and the variance $\sigma^2$ will not be linear in the number of steps $T$. 

The quantum random walk of $T$ steps is defined as the transformation $U^T$, where $U$, acting on ${\cal H}={\cal H}_C \otimes {\cal H}_P$ is given by
\be \label{Eq:U}
U=S \cdot (C \otimes I).
\ee
To illustrate the departure of the quantum walk from its classical counterpart let us evolve the walk (without intermediate measurements) for some steps starting in the initial state $|\Phi_{in}\ra=|\da\ra \otimes |0\ra$ and study the induced probability distribution on the positions.
\begin{eqnarray}\label{Eq:leftwalk}
|\Phi_{in}\ra &\stackrel{U}{\longrightarrow}& \frac{1}{\sqrt{2}}(|\ua\ra \otimes |1\ra - |\da \ra \otimes |-1\ra) \nonumber \\ &\stackrel{U}{\longrightarrow}& \frac{1}{2} (|\ua\ra \otimes |2\ra - (|\ua\ra - |\da \ra) \otimes |0\ra +|\da \ra \otimes |-2\ra) \nonumber \\
 &\stackrel{U}{\longrightarrow}& \frac{1}{2 \sqrt{2}} (|\ua\ra \otimes |3\ra + |\da \ra \otimes |1\ra + |\ua \ra \otimes |-1\ra \nonumber \\ && - 2 |\da \ra \otimes |-1 \ra - |\da \ra \otimes |-3\ra).
\end{eqnarray}
This example already shows that the probability distribution induced by the quantum walk differs from the classical one. In Table \ref{Tab:1} we list the classical distribution and in Table \ref{Tab:2} the corresponding quantum distribution if we measure the position register after $T$ steps.

\begin{figure}[t]
%\begin{table}
 \begin{tabular}{|c|c|c|c|c|c|c|c|c|c|c|c|} 
\hline 
% &---&---&---&---&---&---&---&---&--- \\ \hline
 \begin{picture}(12,12)\put(-1.5,12){\line(1,-1){15}}
    \put(-1.7,-2.5){$T$}\put(8,5){$i$} \end{picture}
 &-$5$&-$4$&-$3$&-$2$&-$1$&$0$&$1$&$2$&$3$&$4$&$5$\\ \hline
0& & & & & &$1$& & & & &\\ \hline
$ 1 $& & & & &$\frac{1}{2}$&&$ \frac{1}{2} $& & & &\\ \hline
$ 2 $& & & &$\frac{1}{4}$& &$\frac{1}{2}$& &$ \frac{1}{4} $& & &\\ \hline
$ 3 $& & &$ \frac{1}{8} $& &$ \frac{3}{8} $&&$ \frac{3}{8} $& &$
 \frac{1}{8} $& &\\ \hline
$ 4 $& &$ \frac{1}{16} $& &$ \frac{1}{4} $&&$\frac{3}{8}$&&$
 \frac{1}{4} $& &$ \frac{1}{16} $&\\ \hline 
5 &$\frac{1}{32}$  &  &$\frac{5}{32}$  &  &$\frac{5}{16}$&&$\frac{5}{16}$& &$\frac{5}{32}$ &&$\frac{1}{32}$\\\hline
\end{tabular}
\smallskip
 \caption{The probability of being at position $i$ after $T$ steps of
 the classical random walk on the line starting in $0$.}
 \label{Tab:1}
%\end{table}
%\begin{table}
 \begin{tabular}{|c|c|c|c|c|c|c|c|c|c|c|c|} \hline 
 \begin{picture}(12,12)\put(-1.5,12){\line(1,-1){15}}
    \put(-1.7,-2.5){$T$}\put(8,5){$i$} \end{picture}
 &$-5$&$ -4 $&$ -3 $&$ -2 $&$ -1 $&$ 0 $&$ 1 $&$ 2 $&$ 3 $&$ 4 $&$5$\\ \hline
 $ 0 $&& & & & &$ 1 $& & & && \\ \hline
 $ 1 $&& & & &$ \frac{1}{2} $&&$ \frac{1}{2} $& & && \\ \hline
 $ 2 $&& & &$ \frac{1}{4} $&&$ \frac{1}{2} $&$ 0 $&$ \frac{1}{4} $& && \\ \hline
 $ 3 $&& &$ \frac{1}{8} $&&$ \frac{5}{8} $&&$ \frac{1}{8} $&&$
 \frac{1}{8} $&& \\ \hline
 $ 4 $&&$ \frac{1}{16} $&&$ \frac{5}{8} $&&$ \frac{1}{8} $&&$
 \frac{1}{8} $&&$ \frac{1}{16} $&\\ \hline 
5 &$\frac{1}{32}$  &  &$\frac{17}{32}$  &  &$\frac{1}{8}$& &$\frac{1}{8}$& &$\frac{5}{32}$ &&$\frac{1}{32}$\\ \hline
 \end{tabular}
\smallskip
 \caption{The probability of being found at position $i$ after $T$ steps of
 the quantum random walk on the line, with the initial state $|\Phi_{in}\ra=|\da\ra \otimes |0\ra$. Note that this distribution starts to differ from the classical distribution from $T=3$ on. Furthermore the quantum random walk is asymmetric with a drift to the left.}
 \label{Tab:2}
%\end{table}
\end{figure}
\begin{figure}[t]
\epsfxsize=8cm
\epsfbox{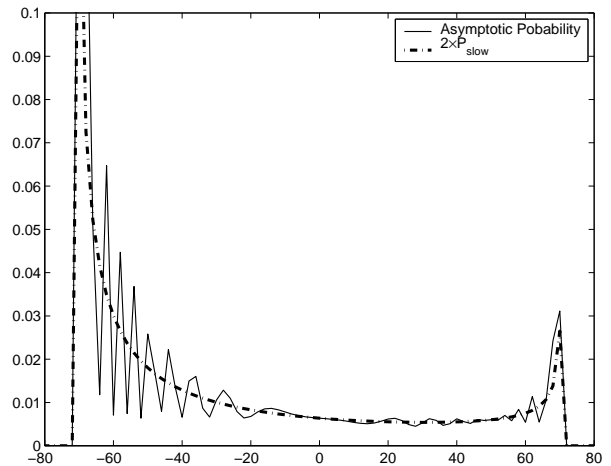}
\caption{The probability distribution of the quantum random walk with Hadamard coin starting in $|\da\ra \otimes |0\ra$ after $T= 100$ steps. Only the probability at the even  points is plotted, since the odd points have probability zero. The dotted line gives a long-wavelength approximation (labeled $P_{slow}$ since it keeps only the slowly varying frequencies \cite{Nayak:00a}), clearly showing the bi-modal character of the distribution.}
\label{Fig:assym}
\end{figure}

Figure \ref{Fig:assym} shows the probability distribution after $T=100$ steps of the quantum walk starting in $|\da\ra \otimes |0\ra$. It is evident that the interference pattern of this walk is much more intricate than the Gaussian obtained in the classical case. One can clearly discern a two-peaked distribution. 

%\begin{center}
%\vbox{
%}
%\end{center} 

A first thing to notice is that the quantum random walk induces an asymmetric probability distribution on the positions, it is 'drifting' to the left. This asymmetry arises from the fact that the Hadamard coin treats the two directions $|\ua\ra$ and $|\da\ra $ differently; it multiplies the phase by $-1$ only in the case of $|\da\ra$. Intuitively this induces more cancellations for paths going right-wards (destructive interference), whereas particles moving to the left interfere constructively.
There are two ways to mend this asymmetry. Inspecting Eq. (\ref{Eq:rightwalk}) which describes the first step of the Hadamard walk starting in $|\ua\ra \otimes |0\ra$ (instead of in $|\Phi_{in}\ra = |\da\ra \otimes |0\ra$ as in Eq. (\ref{Eq:leftwalk})) and iterating the walk in Eq. (\ref{Eq:rightwalk}) more we see that the walk starting in $|\ua\ra \otimes |0\ra$ has a drift to the right side, exactly opposite to the walk in Eq. (\ref{Eq:leftwalk}). To obtain a symmetric distribution we can start the walk in a superposition of $|\ua\ra$ and $|\da\ra$ and make sure that these two drifts do not interfere with each other. This can be achieved starting in the symmetric state $|\Phi_{sym}\ra=\frac{1}{\sqrt{2}}(|\ua\ra +i |\da \ra) \otimes |0\ra$. Since the Hadamard walk does not introduce any complex amplitudes the trajectories from $|\ua\ra$ will stay real and the ones from $|\da\ra$ will be purely imaginary; they will hence not interfere with each other, making the total distribution symmetric. 

Another solution to eliminate asymmetry of the walk is to use a different (balanced) coin, namely
\be
Y=\frac{1}{\sqrt{2}} \left( \begin{array}{cc} 1 & i \\ i & 1 \end{array} \right).
\ee
It is not hard to see that this coin treats $|\ua\ra$ and $|\da\ra$ in the same way and does not bias the walk, independently of its initial coin-state.

\begin{figure}[t]
\epsfxsize=8cm
\epsfbox{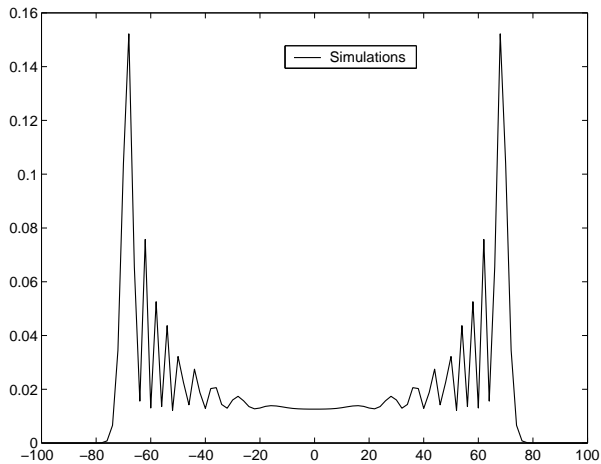}
\caption{The probability distribution obtained
from a computer simulation of the Hadamard walk with a symmetric initial
condition \cite{Nayak:00a}.
The number of steps in the walk was taken to be~$100$. Only the
probability at the {\em even\/} points is plotted, since the odd points
have probability zero.}
\label{Fig:symmetric}
\end{figure}

Figure \ref{Fig:symmetric} shows the probability distribution on the positions of a symmetric quantum walk. 

%\begin{center}
%\vbox{
%}
%\end{center}
The pattern of the probability distribution is very intricate - a signature of the quantum world. The multitude of oscillations makes it hard to analyze the moments of this walk precisely. Using two different approaches - combinatorial techniques (recursions, path counting -  initiated by Meyer \cite{Meyer:96a}) - and a more physical path-integral approach (which expresses the walk in the Fourier domain in terms of integrals that are amenable to asymptotic analysis) Ambainis et al. \cite{Ambainis:01b} give an asymptotic analysis of the variance of the quantum random walk\footnote{See Nayak and Vishwanath \cite{Nayak:00a} for more details on the path-integral approach.}.

As mentioned before the classical symmetric random walk on the line after $T$ steps has a variance $\sigma^2=T$, so the expected distance from the origin is of order $\sigma = \sqrt{T}$. By contrast it can be shown that the quantum random walk has a variance that scales with $\sigma^2 \sim T^2$, which implies that the expected distance from the origin is of order $\sigma \sim T$ - the quantum walk propagates quadratically faster! 

Furthermore the walk spreads roughly uniformly over the positions in the interval $[-\frac{T}{\sqrt{2}},\frac{T}{\sqrt{2}}]$ as can be seen from Fig. \ref{Fig:symmetric} and shown analytically \cite{Ambainis:01b}. This is again in stark contrast to the classical case in which the distribution is peaked around the origin and drops off exponentially several standard deviations $\sigma$ away from the origin. 

To uncover more striking differences of the quantum walk we can place it on a bounded line, either one-sided or two-sided \cite{Ambainis:01b,Bach:02a}. In other words we can insert one or two absorbing boundaries on the line. Formally an absorbing boundary in position $|b\ra$ corresponds to a partial measurement of the process at every time step. More precisely the unitary step $U$ of the walk will be followed by a measurement $M_b$ on the position space ${\cal H}_P$ described by the two projections onto $|b\ra$ and $B_\perp$ (the space orthogonal to $|b\ra$):
\be \label{Eq:Mb}
M_b |\psi\ra = \left\{ \begin{array}{cl} |b\ra & p_b=|\la b|\psi\ra|^2 \\ \frac{|\psi\ra - \la b|\psi\ra |b\ra}{\sqrt{1-|\la b|\psi\ra|^2}} & p_{B_\perp}=1-|\la b|\psi\ra|^2 
\end{array}\right.
\ee
where $p_b$ is the absorption probability and $p_{B_\perp}$ its complement. For example if the current state of the system before measurement is given by 
\be
|\Psi\ra=\frac{1}{\sqrt{14}}(2 |\ua\ra \otimes |0\ra  - |\ua \ra \otimes |1\ra + 3 |\da \ra \otimes |1\ra)
\ee
then after the measurement $M_0$ the state will be $|\ua\ra \otimes |0\ra$ with probability $2/7$ (in which case we say absorption occurred and stop the walk) or otherwise in the state $\frac{1}{\sqrt{10}}(- |\ua \ra \otimes |1\ra + 3 |\da \ra \otimes |1\ra)$ with probability $5/7$.

Now define one step of the random walk with absorbing boundaries as $U$ (Eq. (\ref{Eq:U})) followed by $M_b$ (or $M_b$ and $M_{b'}$ in the case of two absorbing boundaries placed in $|b\ra$ and $|b'\ra$). If the measurement result of $M_b$ (resp. $M_{b'}$) gives $|b\ra$ (resp. $|b'\ra$) the walk is stopped, otherwise the next iteration is applied. 

\begin{figure}[t]
\epsfxsize=8cm
\epsfbox{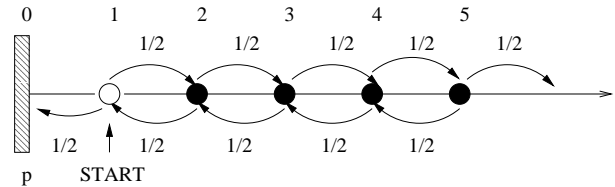}
\caption{The classical symmetric random walk on the line with one absorbing boundary placed in the origin. The starting point of the walk is in position $1$.}
\label{Fig:lineclass}
\end{figure}

Let us review what is known about a classical random walk with one absorbing boundary. For concreteness let us assume that the walk is started in position $1$ and that the boundary is placed in position $b=0$, as in Fig. \ref{Fig:lineclass}.

%\begin{center}
%\vbox{
%}
%\end{center}

It is well known that the probability $p$ to ever get absorbed by the wall in the origin is $p=1$. This is very easy to see via a recursive reasoning on $p$: starting in $1$ the walk hits $0$ with probability $1/2$, in which case it gets absorbed. Otherwise (with probability $1/2$) its position is $2$. The probability $P_{20}$ to ever hit $0$ from $2$ is the probability $p_{21}$ to ever hit $1$ from $2$ times the probability $p_{10}$ to ever hit $0$ from $1$. Both $p_{21}$ and $p_{10}$ are equal to $p$ (the walk is homogeneous in space), which leads to the recursion
\be
p=\frac{1}{2}+\frac{1}{2}p_{21}p_{10}=\frac{1}{2}(1+p^2)
\ee
with the solution $p=1$. In the classical world a random walk always gets absorbed by a boundary and never escapes to infinity\footnote{Note that a classical random walk on the line is {\em recurrent}, i.e. it hits every point infinitely often.}.

Not so for the quantum walk. Careful analysis \cite{Ambainis:01b,Bach:02a} shows that the quantum Hadamard walk with absorbing boundary in $|0\ra$, starting in the state $|\ua\ra \otimes |1\ra$ has a non-vanishing probability to escape. The probability to be absorbed at $|0\ra$ turns out to be $p_{quantum}=2 / \pi \approx 0.63662$, which gives an escape probability of $\approx 0.36338$. This behavior is general: for different initial states and distance to the boundary the numerical value for the absorption probability $p_{quantum}$ changes, but the escape probability is non-zero and finite in all cases.

This is only the tip of the iceberg. After the seminal papers by Ambainis et al. \cite{Ambainis:01b} and Aharonov et al. \cite{Aharonov:01a} a lot of research has been focused on quantum random walks uncovering more and more results of quantum 'weirdness'. We will survey more of them in the following Section.

\begin{figure}[t]
\epsfxsize=7cm
\epsfbox{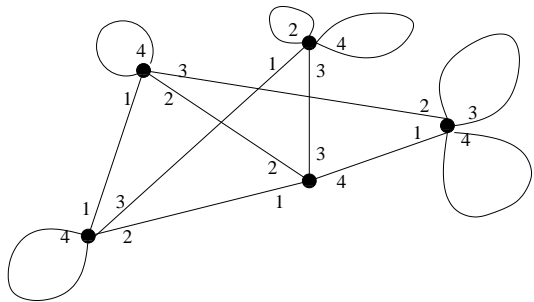}
\caption{A graph with various degrees and a labeling of the edges for each vertex. Note that the label associated with and edge $(v,w)$ can be different from $v$'s end and from $w$'s end. We have added self-loops to the graph to make it regular so that a single $4$-dimensional coin can be used.}
\label{Fig:gengraph}
\end{figure}

In the spirit of the random walk on the line we can generalize the definition to quantum random walks on general graphs. Various aspects of quantum random walks on graphs and in higher dimensions have recently been studied in \cite{Moore:01a,Kempe:02b,Mackay:02a,Yamasaki:02a,Tamon:02a,Shenvi:02b}. To set the stage let us define the random walk for $d$-regular graphs first. These are graphs with vertex-degree $d$, i.e. each vertex has $d$ outgoing edges. The coin Hilbert space ${\cal H}_C$ is of dimension $d$. For every vertex let us internally label all the outgoing edges with $j \in 1 \ldots d$ such that each edge has a distinct label. Let us call $e_v^j$ an edge $e=(v,w)$ which on $v$'s end is labeled by $j$. The state associated with a vertex $v$ pointing along an edge labeled $j$ is $|j\ra \otimes |v\ra$ (corresponding to $|\ua \ra \otimes |i\ra$ on the line or circle). Now we can define a conditional shift operation $S$
\be
S |j \ra \otimes |v\ra =\left\{ \begin{array}{cc} |j\ra \otimes |w\ra & \mbox{if}\, e_v^j=(v,w) \\
0 & \mbox{o/w} \end{array}\right.
\ee
$S$ moves the particle from $v$ to $w$ if the edge $(v,w)$ is labeled by $j$ on $v$'s side. If the graph is not $d$-regular we can still define $S$ in the same way if we let $d$ be the maximum degree of any vertex in the graph. Figure \ref{Fig:gengraph} shows a graph and a valid labeling.

%\begin{center}
%\vbox{
%}
%\end{center}

The coin-flip $C$ is now a $d$-dimensional unitary transformation, which gives a lot of freedom in the choice of the coin-flip operation. The choice of $C$ will have to be guided by the specifics and symmetries of the random walk we wish to obtain. We have already seen that even in the simple case of a walk on the line the choice of the coin operator can have a serious impact on the shape of the resulting distribution, and this effect is only amplified in higher dimensions (see \cite{Mackay:02a} for examples of the effects of various coins on a two- and three-dimensional grid) or on graphs of higher valence (degree). 

If the graph is not regular, i.e. if the vertices have varying degrees, we can use one of several tricks to define a random walk with a coin. Either we add self-loops to each vertex of degree less than $d$ ($d$ is the maximal degree) and include them into the labeling (with the same label on both its ends), as in Fig. \ref{Fig:gengraph}. The shift applied to a self-loop will just keep the walk in place. Another option is to keep the irregularity of the graph and use different coin operators $C_{d'}$ (of different dimensions) for vertices of degree $d' \leq d$. In that case the coin-operation has to be conditioned on the position of the particle and can no longer be written in the separable form of Eq.~(\ref{Eq:U})\footnote{For the use of a discrete quantum walk with a coin that depends on the position register to solve a quantum search problem see \cite{Shenvi:02b}.} as $(C \otimes I)$. However whenever we can define the corresponding classical walk we are able to define the quantum walk and implement it efficiently on a quantum computer (see Sec. \ref{Sec:cs}).

If we wish to retain the property in general graphs that the coin is balanced (i.e. every direction is obtained with equal probability if we measure the coin space) we can use the following coin which generalizes $H$
\begin{equation}
\begin{array}{lcc}
DFT&=&\frac{1}{\sqrt{d}}\left( \begin{array}{ccccc}
1 & 1 & 1 & \ldots & 1\\
1 & \omega & \omega^2 & \ldots & \omega^{d-1}\\
&& \ldots & \\
1 & \omega^{d-1} & \omega^{2(d-1)} & \ldots & \omega^{(d-1)(d-1)}
\end{array} \right).
\end{array}
\end{equation}
Here $\omega=\exp(2 \pi i/d)$ is a $d$-th root of unity. Clearly the unitary $DFT$-coin\footnote{The name of the coin '$DFT$' comes from 'Discrete Fourier Transform'. This transformation is a cornerstone in quantum algorithms and a building block for algorithms like factoring. It has the property that it is efficiently implementable on a quantum computer. We will specify this statement more in Sec. \ref{Sec:cs}.} transforms each direction into an equally weighted superposition of directions such that after measurement each of them is equally likely to be obtained (with probability $1/d$). 

\begin{figure}[t]
\epsfxsize=7cm
\epsfbox{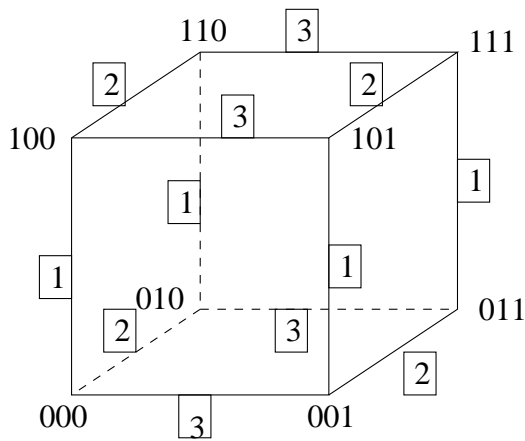}
\smallskip
\caption{The hypercube in $d=3$ dimensions. Vertices correspond to $3$-bit strings. Edges are labeled by $1,2,3$ (boxed) according to which bit needs to be flipped to get from one vertex of the edge to the other.}
\label{Fig:hyper}
\end{figure}

Let us give an example where a non-balanced coin is used. The graph is the hypercube of dimension $d$. Figure \ref{Fig:hyper} shows the hypercube in $3$ dimensions. The vertices of the $d$-dimensional hypercube can be enumerated by $d$-bit strings of $0$'s and $1$'s. Two vertices given by two bit-strings are connected whenever they differ by exactly one bit (e.g. $001101$ and $011101$). If we define the {\em Hamming-distance} $d_H(x,y)$ between two bit-strings $x$ and $y$ to be the minimum number of bits that need to be flipped to obtain $y$ from $x$; then the hypercube connects all vertices of Hamming distance $d_H=1$. The number of $1$'s in a bit-string is called its {\em Hamming weight}.

%\begin{center}
%\vbox{
%}
%\end{center}

Note that the labeling of the edges of the hypercube can be chosen such that for each edge the labels from both sides coincide\footnote{Such a labeling allows to color the edges of a graph (with colors $1,2,\ldots,d$) such that the colors of the incident edges of a vertex are all different. It is a well known fact from graph theory that every graph of maximum degree $d$ can be consistently edge-colored with at most $d+1$ colors. In the case of the $d$-dimensional hypercube only $d$ colors are needed. }. 

\begin{figure}[t]
\epsfxsize=7cm
\epsfbox{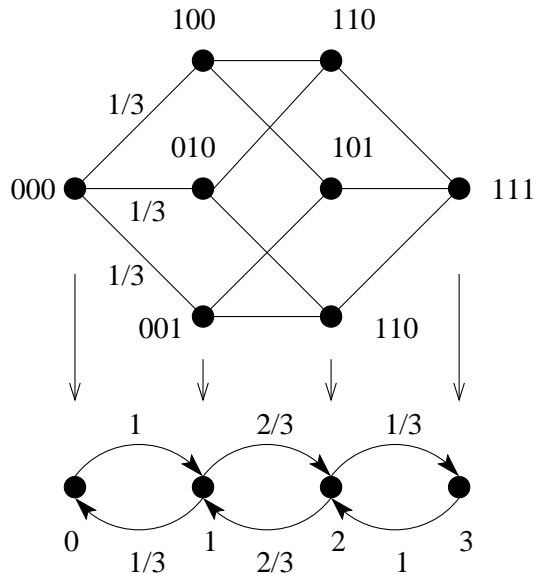}
\smallskip
\caption{The classical simple random walk on the $3$-dimensional hypercube reduced to the walk on a line of $4$ points. In the original walk the transition probabilities are $1/3$. If we bundle nodes of same Hamming weight $w_H$ to a single point on the line we obtain a biased random walk on the line. The new transition probabilities for the walk on the line are indicated. For the walk on the $d$-dimensional hypercube we can reduce the graph to a line of $d+1$ vertices. The transition probability from vertex $i$ to vertex $i+1$  in the general case is given by $p_{i,i+1}=\frac{d-i}{d}$ and from $i+1$ to $i$ by $p_{i+1,i}=\frac{i+1}{d}$.}
\label{Fig:collapse}
\end{figure}

The classical simple random walk on the hypercube (where in each step a neighboring vertex is chosen with probability $1/d$) has a high symmetry. If we start the walk in the vertex $00 \ldots 0$, for example, then all $d$ vertices of Hamming weight $1$ can be interchanged without modifying the random walk. Similarly all vertices of equal distance to the starting vertex, i.e. of equal Hamming weight, can be interchanged without changing the walk. This implies that all vertices of same Hamming weight have the same weight in the probability distribution of the random walk. This allows us to 'cumulate' all the vertices of same Hamming weight into a single vertex and to reduce the symmetric random walk on the hypercube to a biased random walk on the line, as is shown in Figure \ref{Fig:collapse} for the hypercube of dimension $3$. Such a reduction to the line will substantially simplify the analysis of the random walk when we are interested in questions like the speed of propagation of the random walk to his opposite corner. 

%\begin{center}
%\vbox{
%}
%\end{center}

If we wish to retain this symmetry property for the quantum random walk it can be shown \cite{Moore:01a} that the only coins that are unitary and permutation symmetric are of the following form
\be
\begin{array}{lcc} \label{Eq:grover}
G_{a,b}&=&\left( \begin{array}{cccccc}
a&b&b&\ldots &b&b\\
b&a&b& \ldots &b&b\\
b&b&a&\ldots &b&b\\
 &&& \ldots &\\
b&b&b&\ldots & a&b\\
b&b&b&\ldots &b& a \end{array} \right)
\end{array}
\ee
where $a$ and $b$ are real, $1-\frac{2}{d} \leq |a| \leq 1$ and $b=\pm (1-a)$. Among all these coins we will pick the coin $G$ with $a=\frac{2}{d}-1$ and $b=\frac{2}{d}$ and will sometimes call it 'Grover-coin'\footnote{We have called the coin '$G$' because the unitary transformation of this coin corresponds to another important transformation in quantum computation: it is often called 'Grover'-operation. It was first introduced by Grover in 1996 \cite{Grover:96a} in his quantum algorithm to search an unordered database. Grover's algorithm is quadratically faster than any classical algorithm, even if it is probabilistic. Grover's operation became another crucial building block for quantum algorithms. } . Among all coins $G_{a,b}$ this is the coin which is the farthest away from the identity operator ($G_{1,0}=I$). $G$ is not a balanced coin because the probability to not change directions ($p=(1-\frac{2}{d})^2$) is different from the probability to flip to one of the $d-1$ other directions ($p=\frac{4}{d^2}$). If we start the quantum walk in an equal superposition of all the directions and measure the coin space ${\cal H}_C$ after each iteration the resulting (classical) walk will have a higher propensity to go back and forth on the same edge than to switch directions. However the permutation invariance will allow us to reduce the quantum walk on the hypercube to a quantum walk on the line. This would not be possible if we used the $d$-dimensional $DFT$-coin, for example. We will return to this graph in Sec. \ref{Sec:cs}.

\subsection{Continuous time quantum random walk}\label{Sec:cont}

The discrete time model of quantum random walks is only one way to intersperse quantum effects with random walks. Another route to generalize the phenomenon described in Section \ref{Sec:gentle} has been taken by Farhi et al. in 1998 \cite{Farhi:98a,Childs:01a}. At the outset this approach seems very far from the discrete time random walk we have described above. Later, similarities will appear, however.

The continuous time random walk takes place entirely in the position space ${\cal H}_P$. No coin space is needed and no coin is flipped. The intuition behind this model comes from continuous time classical Markov chains. Let us illustrate this with the simple classical random walk on a graph with vertex set $V$. A step in the classical random walk can be described by a matrix $M$ which transforms the probability distribution over $V$. The entries $M_{i,j}$ give the probability to go from $i$ to $j$ in one step of the walk. Let $\vec{p}^t=(p^t_1,p^t_2,\ldots,p^t_{|V|})$ be the probability distribution over the vertices of $V$ at time $T$. Then
\be \label{Eq:discretejump}
p_i^{t+1}=\sum_j M_{i,j} p^t_j
\ee
which means 
\be \label{Eq:dj2}
\vec{p}_{t+1}=M\vec{p}_t.
\ee
The entries of $M_{ij}$ are non-zero only if there is a non-zero probability to go from $i$ to $j$, i.e. when $i$ and $j$ are connected. The entry $M_{ij}$ is the probability to go from $i$ to $j$ and is equal to $\frac{1}{d_i}$ (in the so called 'simple' random walk) where $d_i$ is the degree of $i$. For instance the matrix $M$ for a simple random walk on a circle of $N$ vertices is given by
\be
\begin{array}{lcc}
M&=& \frac{1}{2}\left( \begin{array}{cccccc}
0&1&0&\ldots & 0 & 1\\
1&0&1&0&\ldots&0\\
0&1&0&1&\ldots&0\\
&&\ldots &&&\\
0&\ldots&1&0&1&0\\
0&\ldots&0&1&0&1\\
1&0&\ldots&0&1&0
\end{array} \right).
\end{array}
\ee
It transforms an initial state $\vec{p}^0=(1,0,\ldots,0)$, corresponding to a starting position of $0$, to $\vec{p}^1=(0,\frac{1}{2},\ldots,0,\frac{1}{2})$, $\vec{p}^2=(\frac{1}{2},0,\frac{1}{4},0,\ldots,0,\frac{1}{4},0)$, $\vec{p}^3=(0, \frac{3}{8},0,\frac{1}{8},0,\ldots,0,\frac{1}{8},0,\frac{3}{8})$ etc. corresponding to the distributions as listed in Table \ref{Tab:1}. 

The process given by the iterations of $M$ transforms the state at integer times only. To make the process continuous in time we can assume that transitions can occur at all times and the jumping rate from a vertex to its neighbor is given by $\gamma$, a fixed, time-independent constant. This means that transitions between neighboring nodes occur with probability $\gamma$ per unit time. The infinitesimal generator matrix $H$ of such a process  is given by
\be \label{Eq:Hamiltonian}
H_{i,j}=\left\{
\begin{array}{cl} 
-\gamma &  i \neq j \mbox{ and }i\mbox{ and }j\mbox{ connected} \\
0 & i \neq j \mbox{ and }i\mbox{ and }j\mbox{ not connected} \\
d_i \gamma & i=j
\end{array} \right.
\ee
If $p_i(t)$ denotes the probability to be at vertex $i$ at time $t$ then the transitions in analogy to Eq. (\ref{Eq:discretejump}) can be described by a differential equation
\be
\frac{d p_i(t)}{dt}=-\sum_j H_{i,j}p_j(t).
\ee
Solving the equation we obtain $\vec{p}(t)=\exp(-Ht)\vec{p}(0)$ in analogy to Eq. (\ref{Eq:dj2}). Note the similar structure of $H$ and $M$. In the theory of classical Markov chains many connections between the discrete and continuous time model can be made and many of the quantities of interest, like mixing times (see Sec. \ref{Sec:markov}) and absorption probability, have similar behavior in both cases \cite{Aldous:notes}.

The idea of Farhi and Gutmann \cite{Farhi:98a} was to carry this construction over to the quantum case. Their key idea is that the generator matrix will become the {\em Hamiltonian} of the process generating an evolution $U(t)$ as
\be
U(t)=\exp(-iHt).
\ee
If we start in some initial state $|\Psi_{in}\ra$, evolve it under $U$ for a time $T$ and measure the positions of the resulting state we obtain a probability distribution over the vertices of the graph as before. 

\begin{figure}[t]
\epsfxsize=7cm
\epsfbox{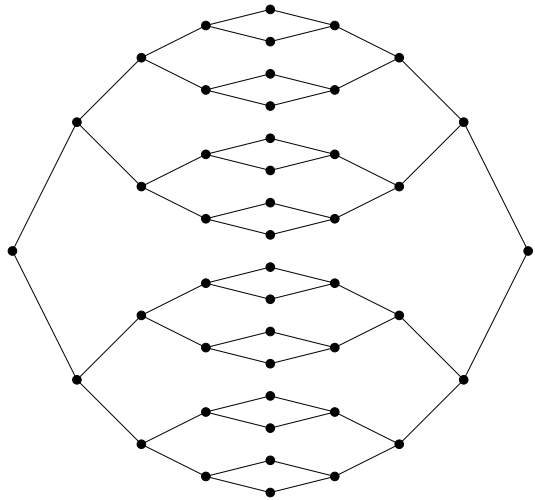}
\caption{The graph $G_4$. Two binary trees of $n=4$ levels are glued together at their leafs in a symmetric fashion \cite{Childs:01a}.}
\label{Fig:Farhigraph}
\end{figure}

With this definitions in place Farhi and Gutmann \cite{Farhi:98a} study the penetration of graphs by the quantum random walk.  In \cite{Childs:01a}  a finite graph is given where classical and quantum walks give an exponential separation in {\em expected hitting time}\footnote{In \cite{Farhi:98a} Farhi and Gutmann already found an example of an infinite tree with a certain leaf and a starting vertex such that the classical random walk takes exponential time to reach the leaf with appreciable probability from the starting vertex whereas the quantum walk hits the leaf very rapidly (in polynomial time). The constructions and proofs are rather involved.}. We will give this example here to illustrate yet another aspect of the difference of classical and quantum random walks, this time regarding the expected hitting time of certain nodes in a graph. The expected hitting time between two points $S$ and $T$ in a graph is the time it takes the random walk on average to reach $T$ starting at $S$. The graph $G_n$ consists of two $n$-level binary trees glued together at their leafs as seen in Figure \ref{Fig:Farhigraph}.

%\begin{center}
%\vbox{
%}
%\end{center}

We start the random walk at the root of the left tree. Rather than the hitting time we study a slightly modified related quantity: the probability of hitting the 
root of the other (right) tree as a function of time.  In other words, we are interested in how long it takes to propagate
from the leftmost vertex to the rightmost vertex as a function of $n$.

Consider the classical random walk first.  The vertices of $G_n$ can be grouped in
columns indexed by $j \in \{0, 1, \ldots, 2n\}$.  Column $0$ contains the
root of the left tree, column $1$ contains the two vertices connected to
that root, etc.  Note that column $n$ contains the $2^n$ vertices in the
middle of the graph and column $2n$ is the root at the right. 

As before in the example of the hypercube (Fig. \ref{Fig:collapse}) the symmetry of the walk allows us to reduce it to a random walk on a line. We need
only keep track of the probabilities of being in the columns.  In the left
tree, for $0<j<n$, the probability of stepping from column $j$ to column
$j+1$ is twice as great as the probability of stepping from column $j$ to
column $j-1$.  However, in the right tree, for $n<j<2n$, the probability of
stepping from column $j$ to column $j+1$ is half as great as the probability
of stepping from column $j$ to column $j-1$.  This means that if you start
at the left root, you quickly move to the middle of the graph, but then it
takes a time exponential in $n$ to reach your destination.  More precisely,
starting in column $0$, the probability of being in column $2n$ after any
number of steps is less than $2^{-n}$.  This implies that the probability of
reaching column $2n$ in a time that is polynomial in $n$ must be
exponentially small as a function of $n$.

We now analyze the quantum walk on $G_n$ \cite{Childs:01a} starting in the state corresponding
to the left root and evolving with the Hamiltonian given by (\ref{Eq:Hamiltonian}).
With this initial state, the symmetries of $H$ keep the evolution in a
$(2n+1)$-dimensional subspace of the $(2^{n+1}+2^n-2)$-dimensional Hilbert
space.  This subspace is spanned by states $|\tilde{j}\ra$ (where $0
\le \tilde{j} \le 2n$), the uniform superposition over all vertices in column $j$,
that is,
\be
  |\tilde{j}\ra 
    = \frac{1}{\sqrt{N_j}} \sum_{a \in \mbox{column} j} |a\ra
\,,
\ee
where
\be
  N_j = \left\{ \begin{array}{ll}
    2^j      & 0 \le j \le n \\
    2^{2n-j} & n \le j \le 2n \,.
    \end{array} \right.
\ee

\begin{figure}[t]
\epsfxsize=7cm
\epsfbox{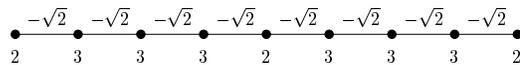}
\caption{The elements of the Hamiltonian of the quantum random walk on $G_n$ when reduced to the line for $n=4$ \cite{Childs:01a}. For convenience $\gamma$ is set to $1$.}
\label{Fig:Farhiline}
\end{figure}

In this basis, the non-zero matrix elements of $H$ are
\begin{eqnarray}
  \la \tilde{j}|H|\tilde{j}\pm 1\ra &=& -\sqrt{2}\gamma \\
  \la \tilde{j}|H|\tilde{j}\ra &=& \left\{ \begin{array}{ll}
    2\gamma & j=0,n,2n \\
    3\gamma & {\mathrm otherwise,} \\
    \end{array} \right.
\end{eqnarray}
which is depicted in Figure \ref{Fig:Farhiline} (for $n=4$) as a quantum random
walk on a line with $2n+1$ vertices.

%\begin{center}
%\vbox{
%}
%\end{center}

\begin{figure}[t]
\epsfxsize=7cm
\epsfbox{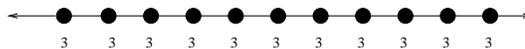}
\caption{Approximation of the finite line by an infinite homogeneous line \cite{Childs:01a}. The diagonal elements of $H$ are all set equal to $3$. The off diagonal elements connecting the vertices are still $-\sqrt{2}$.}
\smallskip
\label{Fig:infline}
\end{figure}

As a first attempt to solve this reduced problem we can approximate the walk on this finite line by a walk on an infinite and furthermore homogeneous line by extending the Hamiltonian and replacing all diagonal entries of $2$ by $3$ as seen in Figure \ref{Fig:infline}.

%\begin{center}
%\vbox{
%}
%\end{center}

 Solving this modified problem is the lore of a quantum physicist and the solution can readily be given in terms of Bessel functions (see \cite{Childs:02a} for a much more complete treatment). It can be seen that the speed of propagation of the random walk on the infinite homogeneous line is linear in the time $T$. In other words there are constants $a < b$ such that when $T \in [an,bn]$ then the probability to measure the walk a distance $2n+1$ from the staring point (i.e. at the root of the right tree starting from the root of the left tree) is of order $1/\sqrt{n}$. 

If we account for the finiteness of the line we are facing the standard quantum mechanical problem of a particle in a square-well potential, which is still not hard to solve. 
It is not entirely obvious that the finiteness of the line and the 'impurity' at level $n$ do not change this behavior significantly\footnote{It is known, for instance, that random impurities on a line will slow a particle down exponentially - a phenomenon related to Anderson localization \cite{Preskill:private}.}. Numerical studies \cite{Childs:01a} confirm that the linear propagation behavior is not significantly changed. To get a hand on the exact hitting time of this walk very recently \cite{Childs:02a} an upper bound for the propagation time is found. The precise statement is that if a time $T$ is chosen randomly $T \in [0,\frac{n^4}{2 \epsilon}]$ and the walk is run for a time $T$, then the probability to measure the state in the root of the right tree is $p_{hit}>\frac{1}{n}(1-\epsilon)$. For algorithmic purposes we can repeat the random walk a polynomial number of times to almost surely find the particle at the right root at least once. 

We can also implement a discrete time random walk on the graph $G_n$ using the $3$-dimensional Grover coin $G$ (Eq. (\ref{Eq:grover})) at the vertices of degree $3$ and the $2$-dimensional one at vertices of degree 2 (which just flips the two directions). Numerical studies \cite{Shenvi:unp} indicate that this discrete time random walk shows an extremely similar behavior to its continuous counterpart. In particular it is possible to reduce the walk to a discrete time walk on the line on which the walk propagates in linear time. This is just one example of the similarity of both models, of which there are many others.

Before we give more results on the difference in behavior of classical and quantum random walks let us pause for a moment to investigate the importance of random walks for algorithms in computer science.

\section{Random walks in computer science} \label{Sec:cs}

Classical random walks form one of the cornerstones of theoretical
computer science \cite{Papadimitriou:book,Motwani:book}. As algorithmic tools, they have been applied to a
variety of central problems \cite{Sinclair:93a}, such as estimation of the volume of a
convex body \cite{Dyer:91a}, approximation of the permanent of a matrix \cite{Jerrum:01a}, and discovery
of satisfying assignments for Boolean formulae \cite{Schoning:99a}. They provide a general paradigm for sampling 
and exploring an exponentially large set of combinatorial structures
(such as matchings in a graph), by using a sequence of simple, local
transitions \cite{Sinclair:93a}.

The behavior of many an algorithm that use random walks depends on quantities like its mixing time (the time it takes the walk to approach its stationary distribution; this is important if we wish to sample from a set of combinatorial structures according to some distribution cf. the Markov Chain Monte Carlo Method\footnote{For an excellent overview of mixing techniques in the algorithmic setting see \cite{Sinclair:93a}.}) or its expected hitting times between two vertices of the underlying graph. We will illustrate the latter point with two simple examples before we give the more formal definitions.

\subsection{Two examples}

\subsubsection{S-T Connectivity}

Let us illustrate how random walks are harnessed in classical computer science by describing how they help to solve the $s-t$ connectivity problem. Given an undirected graph $G$ with vertex set $V$ of size $|V|$, the problem is to decide whether two vertices $s$ and $t$ are connected or not. This problem is a natural abstraction of a number of graph search problems and is important in the study of space bounded complexity classes in computer science \cite{Motwani:book}.

It is easy to see that a standard graph search algorithm that keeps track of all the vertices the search has visited can do the job in a time linear in the number of edges of the graph. Let us assume now that our computer is limited in memory and can keep track only of a few vertices at a time\footnote{More formally we can define a probabilistic Turing machine with logarithmic space. The class of languages this Turing machine can decide in polynomial time is called $RLP$. Random walk arguments show that $s-t$ connectivity is in $RLP$. For an introduction to complexity and randomized algorithms see  \cite{Motwani:book,Papadimitriou:book}.}. It can perform a random walk on the vertices of the graph starting in $s$. At each step of the walk the computer choses one of the adjacent edges with equal probability and moves to the vertex that is connected to the current position by that edge. If the computer reaches $t$ in the course of the walk it outputs YES. If it never sees $t$ after $T=|V|^3$ steps it outputs NO. 

Clearly the machine will never output YES when $s$ and $t$ are not in the same connected component. It turns out that the simple random walk on a graph with $|V|$ vertices has at least probability $1/2$ to hit any given vertex starting from any other in $|V|^3$ steps. This means that our machine will err with probability $1/2$ when it outputs NO. We can make this probability arbitrarily small by repeating the algorithm several times. Assume that $s$ and $t$ are connected. Then the error probability of an algorithm that iterates the above procedure $k$ times is simply the probability to output NO $k$ times. Since we assume that each iteration is independent of the previous one, this probability is given by $(1/2)^k$. If we want to make this probability very small ($\epsilon$) we need to repeat the algorithm roughly $k=\log(\frac{1}{\epsilon})$ times.

\subsubsection{2-SAT}

Let us give another computational problem that can be solved by random walks on graphs, 2-SAT. Let us define more generally the class of decision problems called SAT. In an instance of SAT we are given a set of logical clauses $C_1,\ldots C_m$. The Boolean inputs are the variables $X_1, X_2, \ldots X_n$ which can take the values $0$ or $1$. They can appear in either uncomplemented ($X_i$) or complemented ($\neg X_i$) form in the clauses. A clause is said to be satisfied if at least one of the variables in it is true (an unnegated variable $X_i$ is true if $X_i=0$, a negated variable $\neg X_i$ is true if $X_i=1$). For example the clause $C=X_1 \vee \neg X_2 \vee X_3$ is not satisfied only if $X_1 X_2 X_3 = 101$. We want to know if all the clauses can be satisfied simultaneously by some assignment of $0$ and $1$ for $X_1, \ldots X_n$ (then we say that the formula $C_1 \wedge C_2 \wedge \ldots \wedge C_m$ is satisfiable and output YES).
 The class 2-SAT is the class of all SAT decision problems where each clause contains only two variables. For example take the following instance of 2-SAT on $X_1,X_2,X_3$:
\begin{eqnarray} \label{Eq:2sat}
\Phi(X_1,X_2,X_3) &= & (X_1 \vee \neg X_2) \wedge (\neg X_1 \vee  X_3) \nonumber \\ &&\wedge (X_2 \vee  X_3) \wedge (\neg X_1 \vee \neg X_3) 
\end{eqnarray}
The underlying decision problem has the answer YES and a satisfying assignment is $X_1 X_2 X_3=110$. 

The class of satisfiability problems is very important in theoretical computer science \cite{Papadimitriou:book}. For instance in a famous piece of work (the Cook-Levin theorem) it has been shown that 3-SAT is as hard as any problem in the class NP. This class contains all problems that can be solved in so called non-deterministic polynomial time. That means that if we guess a solution then we can verify that it is indeed a solution in polynomial time. For instance if we were presented with a guess for the satisfying assignment of a formula in 3-SAT we are able to verify this with a machine running in time polynomial in the number of variables $n$ by checking all the clauses $C_1, \ldots C_m$. A plethora of decision and optimization problems are in the class NP. The Cook-Levin theorem states that if we had an efficient algorithm to solve 3-SAT (i.e. to find an answer YES/NO to the decision problem) we can change this into an algorithm to solve any other problem in NP with only polynomial overhead in time. Among the variety of problems known to be in NP we can find Shortest Common Superstring (relevant for gene-reconstruction from sequencing in biology), the Traveling Salesman Problem (important for tour-optimization), Bin Packing problems, the Minesweeper game and Protein Folding, to name just a few. It is not known if a polynomial time (in $n$) algorithm to solve $3$-SAT exists (the famous $P\stackrel{?}{=}NP$ problem).

\begin{figure}[t]
\epsfxsize=7cm
\epsfbox{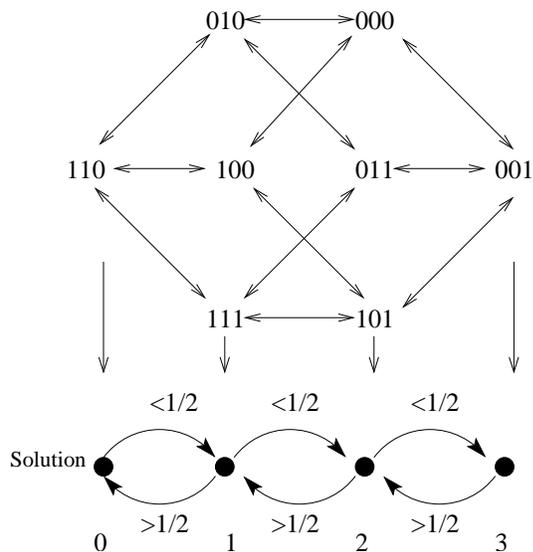}
\smallskip
\caption{A random walk on assignments to the 2-SAT formula of Eq. (\ref{Eq:2sat}). Each step increases or decreases the Hamming distance to the true assignment $110$ by $1$. This gives rise to a random walk on the line.}
\smallskip
\label{Fig:2SAT}
\end{figure}

Let us now give a random walk algorithm that solves 2-SAT in time proportional to $n^2$. To simplify things imagine that our formula is satisfiable and that it has a single truth assignment (in the example of Eq. (\ref{Eq:2sat}) this is $110$). The algorithm picks a random assignment of $0,1$ for the variables $X_1,\ldots,X_n$. It checks the clauses one by one if they are satisfied  until it either runs out of clauses (this means the assignment was satisfying, the algorithm stops and outputs YES) or it finds a clause $C$ that is not satisfied. Say the two variables in the clause are $X_i$ and $X_j$. In the true assignment one of these two variables must be flipped ($0 \leftrightarrow 1$). Now the algorithm choses one of the two variables $X_i$ or $X_j$ randomly and flips it. This is the new assignment and the algorithm restarts anew checking the clauses one by one etc. Let us look what happens to the Hamming distance between the true assignment ($110$ in the example of Eq. (\ref{Eq:2sat})) and the current assignment of the algorithm when we come to flipping $X_i$ or $X_j$. In the true assignment one of the variables $X_i,X_j$ must be flipped (since it satisfies the clause $C$). So with probability $1/2$ the algorithm flips the right variable. In this case the Hamming distance between the new assignment and the true assignment has decreased by $1$. Or the algorithm flips the wrong variable, in which case the Hamming distance increases by $1$\footnote{It can be that flipping either of $X_i$ or $X_j$ can lead toward a true assignment, so the probability to decrease the Hamming distance to some true assignment is at least $1/2$, but can be larger.}. So we are faced with a random walk on the Hamming distances with transition probabilities $1/2$. It is well known that such a walk on the line of length $n$ will hit the corner with high probability after $T \sim n^2$ steps. Figure \ref{Fig:2SAT}  illustrates the algorithm for the example of Eq. (\ref{Eq:2sat}).

%\begin{center}
%\vbox{
%}
%\end{center}

Although this random walk algorithm is not the optimal one to solve 2-SAT it is very instructive. Many randomized algorithms can be viewed as random walks. The efficiency of these random algorithms is then closely related to the expected {\em hitting time} from a vertex (the initial state) to another (the solution). In fact the best known algorithm for 3-SAT \cite{Schoning:99a} relies on a sophisticated version of this random walk idea. 

If a quantum random walk could improve the hitting time on certain graphs this could give rise to efficient quantum algorithms for problems that cannot be solved efficiently on a classical computer. It is yet another incentive to study the behavior of quantum random walks.

\subsection{Classical Random Walks}\label{Sec:markov}

Let us state without proofs some known facts about random walks on graphs \cite{Feller:book}.

A {\em simple random walk} on an undirected graph $G(V,E)$, 
is described by repeated applications of 
 a stochastic matrix $M$, where $M_{i,j}=\frac{1}{d_i}$
if $(i,j)$ is an edge in $G$ and $d_i$ the degree of $i$ (we use the notation introduced in Sec. \ref{Sec:cont}). 
  If $G$ is connected and non-bipartite, then the distribution
 of the random walk, 
$\vec{p}^T=M^T \vec{p}^0$ converges to a {\em stationary 
 distribution $\vec{\pi}$} which is independent of the initial distribution $\vec{p}^0$.
For a $d-$regular $G$ (all nodes have the same degree $d$), 
the limiting probability distribution is uniform over the nodes of the graph. 

In many computational problems the solution can be found if we are able to sample from a set of objects according to some distribution\footnote{See \cite{Sinclair:93a} for an overview of this technique, called Markov Chain Monte Carlo.}. This problem is often approached by setting up a random walk on a graph whose nodes are the objects we want to sample from. The graph and the walk are set up in such a way that the limiting distribution is exactly the distribution we wish to sample from. To sample we now start the walk in some random initial point and let it evolve. This type of algorithm is only efficient if the random walk approaches the limiting distribution {\em fast}.

There are many definitions which capture  the rate of convergence
to the limiting distribution. A survey can be found in \cite{Lovasz:98a}.   
A frequently used quantity is the {\em mixing time} given by  
\be
M_\epsilon=\min \{T| \, \forall t\ge T ,\vec{p}^0 : \,  
||\vec{p}^t-\vec{\pi}|| \leq \epsilon\}, 
\ee
where we use the {\em total variation 
distance} to measure the distance between two distributions $\vec{p},\vec{q}$:
\be 
\|\vec{p}-\vec{q}\|= \sum_i |\vec{p}_i-\vec{q}_i|.
\ee

It turns out that the mixing time is related to the
{\em gap} between the  largest 
eigenvalue $\lambda_1 =1$ of the stochastic matrix $M$,  and the
second largest eigenvalue $\lambda_{2}$ in the following way \cite{Lovasz:98a}
\begin{eqnarray}
\frac{\lambda_2}{(1-\lambda_2) \log 2 \epsilon} &\leq& M_\epsilon \nonumber \\ &\leq& \frac{1}{(1-\lambda_2)} (\max_{i} \log \vec{\pi}_i^{-1}+\log \epsilon^{-1}).
\end{eqnarray}
This powerful theorem provides a very useful connection between mixing times and the second largest eigenvalue $\lambda_2$ of the transition matrix $M$. In particular it shows that $\lambda_2$ is the only eigenvalue of $M$ that really matters to determine the mixing behavior of the walk. This will be very different in the quantum case.

{\em Circle:}
It is well known that for the
 simple random walk on an $N-$circle, the mixing time is quadratic, 
 $M_\epsilon \sim N^2\cdot \log(1/\epsilon)$. Similarly the expected time $T$ we need to run the walk such that starting at a vertex $i$ we hit a certain vertex $j$ with probability close to $1$ is $T \sim N^2$.

{\em Hypercube:} For the $d$-dimensional hypercube defined in Sec. \ref{Sec:discrete} the mixing time scales with $d$ as $M_\epsilon \sim d \log d \log (1/\epsilon)$. The expected hitting time $T$ from one corner of the hypercube, say $00\ldots 0$, to its opposite, $11 \ldots 1$, scales exponentially with the dimension $d$ as $T \sim 2^d$. We will see a sharp quantum improvement here.

\subsection{Quantum Computers and Circuits}

We have developed an intuition for why quantum random walks could be useful for algorithms. Before we describe some more results on the mixing and hitting times of quantum random walks let us first expand a little on the {\em machine} that would implement the quantum random walk - the {\em quantum computer}. We will only give a rushed collection of facts to acquaint the reader with the basics, for a very detailed introduction see \cite{Nielsen:book}.

A quantum computer functions according to the laws of quantum mechanics. In particular its 'computations' have to be {\em unitary} transformations on the state space. 

{\em Qubits:} A classical computer operates on strings of bits. Similarly a quantum computer acts on {\em qubits}. Each qubit is a two level system spanned by the two states $\{|0\ra,|1\ra\}$ (or $\{|\ua \ra,|\da \ra\}$ to make a closer connection to spins). We can think of the state Hilbert space of $n$ qubits to be the tensor product of $n$ $2$-dimensional qubit spaces ${\cal H}_2^{\otimes n}$. Another way to think of the states is to say that they are spanned by $n$-bit strings $|x_1 x_2 \ldots x_n \ra$ with $x_i \in \{0,1\}$. A state of the quantum computer can be in any superposition of these basis states. For a physicist it might be easier to think about a collection of $n$ spin-$\frac{1}{2}$ particles. 

{\em Gates:} The computation of a classical computer can be described by a circuit consisting of gates (like AND or NOT) that act on one or two bits. The computer is initialized in some input state and computes some Boolean function of it. Similarly a quantum computer can be modeled by a circuit consisting of unitary gates that act on one or two qubits only. The goal is to compute some unitary transformation on the initial state.

It is well known that every Boolean function can be decomposed into a succession of gates from a universal set (like AND and NOT). In other words any computation on a classical computer can be written as a succession of AND and NOT. One of the achievements of quantum computing theory has been the insight that any unitary computation on qubits can be decomposed into a succession of quantum gates that act on one or two qubits only. One such universal quantum gate set is given by single qubit unitary matrices $U_{single}$ (these are $2$ by $2$ matrices giving all the generalized rotations of a single spin-$\frac{1}{2}$ system) together with the so called CNOT
\be
CNOT = \left( \begin{array}{cccc} 1 & 0 & 0 & 0 \\0 & 1 & 0 & 0 \\ 0 & 0 & 0 & 1\\0 & 0 & 1 & 0 \end{array} \right).
\ee
To build a quantum computer then means to build a device that can implement single qubit unitaries together with a two-body transformation that implements the CNOT\footnote{In fact any  genuine two-body interaction is sufficient.}.

The question is whether such a quantum computer with its unitaries can even perform the computations of a classical computer. After all, unitary transformations are reversible, so all the quantum computations are reversible and the class of these reversible computations might not contain some classical circuits. It has been shown by Bennett \cite{Bennett:73a} that this is not a serious problem: Any classical circuit can be made reversible by replacing its gates by a reversible three bit gate and adding some more scratch space. Furthermore these reversible gates can be implemented unitarily by a quantum computer. In particular this implies that a quantum computer can emulate any classical computation efficiently: quantum computers are at least as strong as their classical brothers. 

The added bonus of using the laws of quantum mechanics is that the unitary power might be strictly larger than the classical one. The quest for quantum algorithms is exactly to explore this added power. 

Returning to random walks we can see that a quantum computer can efficiently implement any discrete time quantum random walk whenever a classical computer can implement the underlying classical random walk. This is because we can take the classical circuit that implements the classical random walk, turn it into a quantum circuit (after making it reversible) and replacing the classical coin-flip by a quantum coin-flip, which can be done efficiently on the small space of coin-qubits. That means that if we take a classical random walk at the basis of some classical algorithm and show that the quantum random walk improves its behavior, we will immediately have a quantum algorithm with some speed-up over the classical. 

The situation is not quite as easy for the model of continuous time quantum random walks. It is not a priori clear how to adapt its continuous evolution to the discrete quantum circuit model. However it has been shown recently \cite{Aharonov:unp,Aharonov:03a} that simulation of continuous time random walks can be done efficiently on a quantum computer.

With this in place we are now set to explore some more differences between classical and quantum random walks.

\subsection{Results from Quantum Random Walks}

\subsubsection{The circle}

The first and maybe easiest finite graph to explore with quantum random walks is the circle of $N$ vertices \cite{Aharonov:01a}. This example has most of the features of walks on general graphs. 

Recall that any classical random walk approaches a stationary distribution independent of its initial state (Sec. \ref{Sec:markov}) - the classical random walk looses its memory. This cannot be true for the quantum random walk. All transformations are unitary and hence reversible, therefore the walk can never lose its recollection of the initial state and therefore it does not converge to a stationary distribution. To be able to speak of mixing toward a stationary distribution we have to introduce some 'forgetting' into the definitions. There is a standard way to do this, the so called {\em Cesaro limit}. Instead of studying the probability distribution $\vec{p}^t$ induced by the random walk after a measurement at time $t$ and its limit as $t$ grows we pick a time $s$, $0 < s \leq t$, uniformly at random, i.e. our probability distribution $\vec{c}^t$ is an average distribution over all measurement results between $1$ and $t$:
\be
\vec{c}^t = \frac{1}{t} \sum_{s=1}^t \vec{p}^s.
\ee 
With these definitions in place it is not very hard to see that $\vec{c}^t$ converges to a stationary distribution. We will show this below for the Hadamard walk on the circle \cite{Aharonov:01a} - also to give an idea about the kind of calculations involved in these types of problems - but the result is true for all finite graphs.

In order to analyze the behavior of the quantum walk, we follow its wave-like patterns using the eigenvectors and eigenvalues of the unitary evolution $U_t$. To observe its classical behavior, we collapse the wave vector at time $t$, $|\Psi_t\ra=U^t|\Psi_0\ra$, into a probability vector $\vec{p}^t$. The probability to measure the particle in position $i$ at time $t$, $\vec{p}^t_i=|\la \ua , i|\Psi_t\ra|^2+|\la \da ,i|\Psi_t\ra|^2$, where we write $|\ua , i\ra$ short for $|\ua \ra \otimes |i\ra$ and similarly for $|\da ,i\ra$. Let $\{(\lambda_k, |v_k\ra) :k=1 \ldots 2N \}$ be the eigenvalues and eigenvectors of $U$. Note that $\lambda_k^*=\lambda_k^{-1}$ because $U$ is unitary. Expanding the initial state $|\Psi_0\ra = \sum_{k=1}^{2N} a_k |v_k\ra$ we obtain $|\Psi_t\ra = U^t |\Psi_0 \ra = \sum_{k=1}^{2N} a_k \lambda_k^t |v_k \ra$. Putting this all together we get for the $i$th component
\be \label{Eq:ct}
\vec{c}^t_i = \frac{1}{t}\sum_{s=1}^t \sum_{\alpha= \ua, \da} \sum_{k,l=1}^{2N} a_k a_l^* (\lambda_k \lambda_l^*)^s \la \alpha ,i|v_k\ra \la v_l|\alpha ,i \ra
.
\ee 
In the limit $t \rightarrow \infty$ we have 
\be
\frac{1}{t} \sum_{s=1}^t (\lambda_k \lambda_l^*)^s \rightarrow \left\{ \begin{array}{cc} 1 & \lambda_k = \lambda_l \\ \lim_{t \rightarrow \infty} \frac{1}{t (1-(\lambda_k \lambda_l^*))}=0 & \lambda_k \neq \lambda_l \end{array}\right.,
\ee
hence 
\be \label{Eq:stat}
\vec{c}^t_i \stackrel{t \rightarrow \infty}{\longrightarrow} \sum_{\alpha= \ua, \da} \sum_{\begin{array}{c} k,l=1\\ \lambda_k =\lambda_l \end{array}}^{2N} a_k a_l^* \la \alpha ,i|v_k\ra \la v_l|\alpha ,i \ra =:\vec{\pi}_i .
\ee
In the case of non-degenerate eigenvalues $\lambda_k$, Eq. (\ref{Eq:stat}) simplifies further to $\vec{\pi}_i=\sum_{\alpha= \ua, \da} \sum_{k=1}^{2N} |a_k|^2 |\la \alpha ,i | v_k \ra |^2 $. 

Eqs. (\ref{Eq:ct}) and (\ref{Eq:stat}) already reveal some crucial differences between classical and quantum walks. Whereas in the classical case the stationary distribution $\vec{\pi}_i$ is independent of the initial state of the walk, in the quantum walk this is not the case in general, as is manifested by the presence of the expansion coefficients $a_k$ in the expression for $\vec{\pi}_i$. However in the case of the circle (and many other graphs from a family called Cayley graphs\footnote{Cayley graphs are graphs corresponding to Abelian groups: The group is given by a set of generators. Each node of the graph represents an element of the group and two nodes are connected if the corresponding elements can be obtained from each other by applying one of the generators.}) the stationary distribution is uniform and independent from the staring state as long as this state is localized in one position \cite{Aharonov:01a}.

Another remarkable difference is the dependence of the mixing time - the rate of convergence to $\vec{\pi}$ - on {\em all} eigenvalues of $U$. Remember that in the classical case the mixing time was governed only by the second largest eigenvalue of the transition matrix $M$ (see Sec. \ref{Sec:markov}). As can be seen from Eq. (\ref{Eq:ct}) the rate of convergence to $\vec{\pi}$ in the quantum case is governed by the terms
\be
\frac{1}{t} \sum_{s=1}^t (\lambda_k \lambda_l^*)^s = \frac{1- (\lambda_k \lambda_l^*)^t}{t(1- (\lambda_k \lambda_l^*))} \leq \frac{1}{t|\lambda_k - \lambda_l|}
\ee
All of these sums over pairs of eigenvalues enter Eq. (\ref{Eq:ct}), and so the mixing time is determined by all $|\lambda_k-\lambda_l|$. 

It is possible to bound the mixing time $M_\epsilon$ of the quantum random walk on the circle as $M_\epsilon \leq \frac{N \log N}{\epsilon ^3}$ with some more lines of algebra \cite{Aharonov:01a}. This gives a nearly quadratic speed-up over the classical walk on the circle, which mixes in time proportional to $N^2$ (see Sec. \ref{Sec:markov})! We have found yet another 'quantum advantage': any algorithm that uses a random walk on a circle can be made quadratically faster on a quantum computer! 

We might ask if there are graphs for which we can achieve even more drastic speed-ups. In computer science the transition between an easy and a hard problem is made between polynomial time and exponential time. Problems that take exponential time to solve are hard (the resources to solve them do not scale well with their size), problems that take only polynomial amount of time are 'easy'\footnote{Although it seems a bit impractical to call a problem that requires time $n^{10}$ in the size of the input $n$ 'easy'.}. The interesting question is now: Can we find graphs such that the quantum walk mixes exponentially faster than its classical brother? Unfortunately the answer is 'NO' in all but possibly very contrived cases. As shown by Aharonov et al. \cite{Aharonov:01a} the quantum walk mixes at most quadratically faster\footnote{on graphs of bounded degree}. 

Yet this is not the end of all hopes for exponential speed-up, as we will see in the next section.

\subsubsection{The hypercube}

It is possible to calculate the eigenspectrum of several matrices $U$ corresponding to a random walk on a graph with some coin $C$. Moore and Russell \cite{Moore:01a} analysed the mixing time of the walk on the hypercube (see Sec. \ref{Sec:walk}) along the lines of the previous paragraph, both in the discrete time case as well as in the continuous one. Their results for the mixing time are not very encouraging, though. The hypercube is one of the graphs that mixes fast to the uniform distribution in the classical case already (it belongs to a class of so called expander graphs which all share this property), namely $M_\epsilon \sim d \log d$ (Sec. \ref{Sec:markov}). 

The discrete time quantum walk turns out to do worse: Its mixing time $M_\epsilon$ is at least $d^{3/2}/\epsilon$. Even more dramatically the continuous time quantum walk on the hypercube does not converge to the uniform distribution at all. 

We have mentioned another quantity of crucial interest in the study of random walks - the expected hitting time. We have shown the importance of fast hitting to solve some algorithmic problems in Sec. \ref{Sec:cs}. Furthermore it is here that we may hope to achieve great quantum advantage: the classical random walk on the hypercube takes exponential time to hit its opposite corner. 

In the quantum world  the notion of hitting can be made precise in several ways (see \cite{Kempe:02b} for a systematic approach). In order to know whether a walk hits a certain point measurements need to be made, yet we do not wish to measure the position of the walk completely to not kill its quantum coherences and make it classical. In \cite{Kempe:02b} two ways to define hitting time are presented: either we can wait for a certain time $T$ (which we somehow determine in advance) and then measure the walk. If the probability to be in the node of interest is close to $1$ we can call $T$ a hitting time. To circumvent the problem of knowing $T$ in advance we can alternatively use a definition similar to the absorbing wall described in Sec. \ref{Sec:walk}. At each step a measurement is performed which only determines if the walk is in the node of interest or not ($M_b$ in Eq. (\ref{Eq:Mb}) of Sec. \ref{Sec:walk}). If it is found there the walk is stopped. We will call $T$ a hitting time  if the walk has stopped before time $T$ with high probability. In the classical case all these notions of hitting time from one corner of the hypercube to its opposite are exponential in $d$. Kempe \cite{Kempe:02b}
 shows that the quantum discrete time walk indeed behaves crucially different: The expected hitting time (both notions) from a corner to its opposite turns out to be polynomial in the dimension $d$! This is an exponential separation of the classical and quantum behavior, similar to what we have seen on the graph $G_n$ for the continuous time quantum walk (Sec. \ref{Sec:cont})\footnote{The continuous time quantum walk on the hypercube also hits the opposite corner in polynomial time \cite{Kempe:02b}, yet another similarity between the two walks.}. It rests to see how this speed-up can be harnessed algorithmically.

\subsubsection{An oracle separation}

The first successful attempt to make an algorithmic statement using exponentially faster quantum hitting times is given by Childs et al. \cite{Childs:02a}. They provide a so called oracle separation using the example described in Sec. \ref{Sec:cont}. We have seen that the continuous time quantum walk penetrates the glued trees from root to root exponentially faster than the classical walk. This can be made into an exponential separation of classical and quantum query complexity in an oracle problem. This means the following: We are given an oracle (in form of a black box that we cannot open), which we can query with some input to get a specific answer. For instance in database search set-ups the oracle encodes a database with one marked item (like a telephone book where we happen to know one specific phone-number and wish to know who it belongs to). Given the input string $x$ (a possible name) as a query, the oracle will reply $0$ if $x$ is not marked (doesn't belong to the phone number) or $1$ if it is the item we are looking for. Imagine that each such query costs a fixed amount; it is our task then to minimize the number of queries. It has been shown that a quantum computer (who can ask a query 'in superposition') can find the marked item with only $\sqrt{N}$ queries, where $N$ is the number of items, whereas any classical algorithm will require of order $N$ queries. 

\begin{figure}[t]
\epsfxsize=8cm
\epsfbox{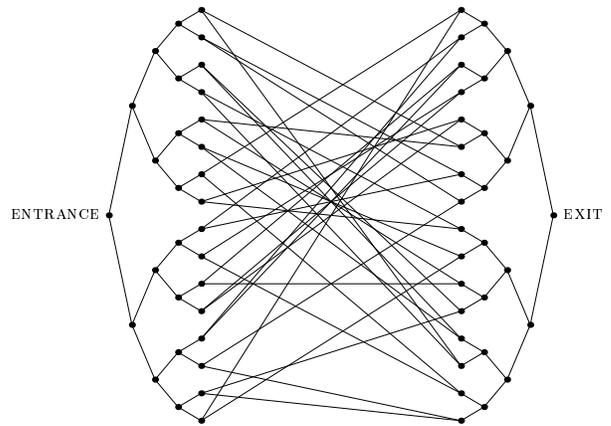}
\caption{The modified graph of Fig. \ref{Fig:Farhigraph}. A big random cycle has been inserted between the two binary trees. This makes the classical task of penetrating the graph from root to root exponentially hard \cite{Childs:02a}.}
\label{Fig:Oraclegraph}
\end{figure}

Childs et al. construct an oracle which encodes a graph similar to the one in Fig. \ref{Fig:Farhigraph}. It turns out that there are classical algorithms (random walks that keep track of the valence of vertices they visit) that can penetrate the graph of Fig. \ref{Fig:Farhigraph} in polynomial time. To make the task classically hard the graph has to be modified by inserting a huge cycle between the trees, as in Fig. \ref{Fig:Oraclegraph}.

%\begin{center}
%\vbox{
%}
%\end{center}
The oracle now encodes the structure of the graph: Every node of the graph has a name unknown to us. We are given the name of the right root and wish to find the name of the left root. Queried with the name of some vertex the oracle outputs the names of the neighbors of this vertex. Classically this oracle has to be queried exponentially many times in $n$ (number of levels of the tree) to obtain the name of the root of the left tree starting with the name of the root of the right tree. Intuitively, the best way to proceed classically is to ask the oracle for the names of the neighbors of the root. With the answer we ask for the names of the neighbors of one of these neighbors and so on making our way through the graph in the form of a random walk. We have seen that it takes exponential in $n$ expected time for the classical random walk to hit the opposite root. Furthermore it can be shown \cite{Childs:02a} that this na\"\i ve way of making our way through the graph is the best a classical algorithm can do. Using the continuous quantum walk\footnote{There are some technicalities involved with using a continuous time walk to make discrete oracle queries \cite{Childs:02a}.} to find the succession of (quantum)-queries to the oracle the problem can be solved with only polynomially many queries using a quantum computer \cite{Childs:02a}\footnote{ Na\"\i vely one can try to obtain a similar oracle separation using the discrete time walk on the hypercube and the speed-up of \cite{Kempe:02b}. However, for the hypercube a fast classical algorithm can be found that performs the task \cite{Childs:02a}.}. For this (unfortunately not very practical) problem we have found an exponential separation between classical and quantum power based on quantum random walks\footnote{Note that this is not the first exponential separation relative to an oracle. The first such separation - based on the quantum Fourier Transform - has been given in 1997 by Bernstein and Vazirani \cite{Bernstein:97a}.}!

\subsubsection{What next?}

We hope that these first algorithms are only the beginning and that more algorithms can be found that employ the power of quantum random walks. For a recent search algorithm based on the discrete time walk on the hypercube see \cite{Shenvi:02b}. 

Many questions are still left open. We will mention some: For instance the quantum random walk on a circle  of even degree (and on many other graphs) does not converge to the uniform distribution (its eigenvalues are degenerate). Rather, its stationary distribution depends on the starting state. It would be interesting to know if one can harness this 'quantum memory' algorithmically. 

Another observation is that the exponential hitting time on both hypercube and glued trees seems to depend crucially on the symmetry of these graphs. Rapid hitting seems to indicate high symmetry and could be used to perform some sort of graph-'tomography', which might have algorithmic use. 

There are a lot of graphs that have not yet been studied in the context of quantum walks (like the walk on the symmetric group, for instance, which is paramount in some interesting algorithmic questions like Graph-Isomorphism).

Furthermore the connection between the two quantum walk models is still not clear. In many cases both walks behave very similarly, in other cases there are differences. It would be useful to make the connection between these two models more precise.

\section{Physical Implementation} \label{Sec:implementation}

We have attempted to convince the reader of the utility to study random walks - both for the interesting physical effects they show as well as for their potential algorithmic use. In both cases it would be interesting to find a physical set-up to implement the quantum random walk. Of course a universal quantum computer can do that - as we have seen in Sec. \ref{Sec:cs}. However, this machine has yet to be built on a large scale. Random walks might exhibit specifics that could ease their implementation in certain physical systems, making use of the structure of the graph underlying the walk or the character of the coin, for example. Such an implementation does not necessarily constitute a full fledged quantum computer but could still be well suited to study random walks or to solve algorithmic problems based on them \cite{Shenvi:02b,Childs:02a}. In this spirit several proposals for various physical systems have been made \cite{Travaglione:02a,Sanders:02a,Duer:02a}. 

As of today it is not clear which physical architecture is the optimal one to build a quantum computer. As we have seen we need qubits and gates which can be implemented in a multitude of ways. This fungibility of quantum information gives rise to proposals ranging from nuclear magnetic resonance systems, optical cavities, solid state architectures, optical lattices to ion traps, to name just a few. Similarly the options to implement a quantum random walk are vast. We will just touch upon a few, without giving a detailed description of the physical system; the reader is referred to the respective references for the details.

Of course the positions of the walk $|i\ra$ need not be real positions in the physical implementation: we can encode them into any discrete degree of freedom. Similarly the coin space need not correspond to the spin of a spin-$\frac{1}{2}$ particle.

For instance Travaglione and Milburn \cite{Travaglione:02a} suggest an {\em ion-trap} based implementation of the random walk on a line. In this set-up an ion is confined in a Radio-Frequency ion trap (see \cite{Sasura:02a,Wineland:98a} for an overview on the ion trap quantum computer). 
The positions $|i\ra$ are encoded into the (discrete) motional states of the ion in the trap. The ion is laser-cooled to its motional ground state $|0\ra$. The electronic internal states of the ion will encode the coin states $|\ua\ra$ and $|\da \ra$. Raman beam pulses are applied to implement a coin-flip. A displacement beam excites the motion correlated to the $|\ua\ra$ internal state of the ion. Flipping the internal states with another Raman pulse and applying the displacement beam again effectively implements the conditional shift. This procedure can be iterated $T$ times. To measure the positions of the walk after $T$ steps it is possible to couple the  motional state to the internal state of the ion and measure the latter (see \cite{Travaglione:02a} for details). The main source of noise in this scheme is the degradation of the internal state of the ion (the coin state). 

With present day technology we might be able to implement a few steps of the random walk\footnote{The major challenge is the preparation of harmonic oscillator states of ionic motion, i.e. the position states.}. If nothing else the variance of the quantum random walk so obtained can serve as a benchmarking protocol for ion trap quantum computers. 

Another recent suggestion by Sanders et al. \cite{Sanders:02a} puts the random walk into a {\em quantum electrodynamics} device. The physical system is an optical cavity traversed by an atom. During the traversal the internal electronic levels of the atom interact with the cavity mode. Furthermore the internal atomic levels can be addressed and manipulated  with a periodic sequence of microwave pulses. The focus of this proposal is to implement the random walk with existing present day technology in a microwave cavity. In this set-up the position of the random walk is encoded into a single cavity mode, whereas the coin states are the internal atomic states\footnote{Aharonov et al. \cite{Aharonov:93a}, when pointing out a first 'quantum random walk effect' in the original 1993 work (see Sec. \ref{Sec:gentle}), already suggested to observe it in the optical cavity setting. The atom in an excited state traverses the cavity. The excited internal atomic state couples to the cavity mode. If upon exit the atom is measured in a rotated basis and a rare event is observed, the wave function of the cavity mode (initially in a coherent state with mean photon number $N$, for instance) can suddenly shift by much more than (classically) expected. As described in Sec. \ref{Sec:gentle} the mean photon number can shift to $N + \delta$ (instead of $N+1$) with $1 \ll \delta$.}. Here a random walk on the circle is achieved. Let us sketch the scheme of \cite{Sanders:02a} in more detail (for the optical background see \cite{Scully:97a}): 

The field mode of the cavity is described by the (infinite) Hilbert space of a harmonic oscillator spanned by photon number states $\{|n\ra: n=0,1,\ldots \}$. Let ${\hat{N}}$ be the number operator acting as $\hat{N}|i\ra=i|i\ra$. For the Hadamard walk on the $N$-circle we will employ a truncated Hilbert space ${\cal H}_N$ spanned by the $N$ states with less than $N$ photons. The random walk will take place on the phase states $\{|\theta_0\ra,|\theta_1\ra, \ldots , |\theta_{N-1}\ra \}$ with
\be
|\theta_j\ra = \frac{1}{\sqrt{N}}\sum_{k=0}^{N-1} e^{\frac{2 \pi ij}{N}k}|k\ra.
\ee 
A shift on the phase states can be induced by the operator $F_\pm=\exp(\pm \frac{2 \pi i}{N} {\hat N})$, i.e. $F_\pm |\theta_j\ra = |\theta_{j \pm 1}\ra$. As in Sec. \ref{Sec:gentle} let the two atomic states we employ be $|\ua \ra$ and $|\da \ra$, and let $S_z$ be as in Eq. (\ref{Eq:Sz}). Then the conditional shift $S$ is given by $S=\exp(-2i S_z \otimes \frac{2 \pi }{N} {\hat N})$ very similar to Eq. (\ref{Eq:shift}). To implement $S$ we use a two-level model including ac-Stark shifts. The atomic levels are highly detuned from the cavity field, and the Hamiltonian for this effect is given by $H=\hbar \chi {\hat N} \otimes 2 S_z$. The Hadamard transformation for the coin-flip on the atomic levels can be realized by a $\pi /2$ pulse. The time $\tau$ between successive pulses is set to be $\tau = 2 \pi i/ \hbar \chi  N$. This implements the random walk.

The standard initial condition would be to have the field in the state that projects to $|\theta_0\ra$ in the truncated space. This is unfeasible experimentally at this point.
To make the random walk readily implementable with present state technology the initial state of the cavity is taken to be a coherent state $|\alpha \ra$ ($\alpha$ real and positive). It is possible to find such states $|\alpha \ra$ that have high overlap with $|\theta_0\ra$ and very small overlap with $|\theta_j \ra$ for $j \neq 0$, at least for $N$ not too large \cite{Sanders:02a}.

Measuring the position of the walker corresponds to detecting a phase shift of an initial coherent cavity field. This can be achieved using homodyne detection (see \cite{Scully:97a} for definitions). All ingredients employed in such an implementation of the random walk are achievable with current technology.

Another upshot of this proposal is that controlled decoherence can be introduced into the cavity. This allows to study the decrease in phase diffusion (transition from a linear spread $\sigma \sim T$ in the quantum case to a classical spread $\sigma \sim \sqrt{T}$)  with noise (see also Sec. \ref{Sec:decoherence}). Note how this decrease in dissipation contrasts sharply with the intuition provided by the fluctuation-dissipation theorem: that the introduction of noise usually results in increased dissipation!

An implementation of random walks in {\em optical lattices} has been proposed by D\"ur et al. \cite{Duer:02a}. This proposal uses neutral atoms trapped in an optical lattice. The walk (on line or circle) takes place in position space given by the minima of an optical lattice. 

An optical lattice consists of periodic optical potentials formed by the standing waves resulting from two counter-propagating waves from two  lasers. Here two identical one-dimensional optical lattices will be used, each of them trapping one of the internal states $|\ua\ra,|\da\ra$ of a neutral atom. The internal states can be hyperfine states of the nuclear spin of the atom. The relative spacing of the potential minima of the optical lattices can be shifted by modulating the angle between the two counter-propagating waves, in particular the minima of the two lattices coincide when the angle is a multiple of $\pi /2 $.
To implement the conditional shift we modulate the angle such that the lattice trapping the $|\ua\ra$-state moves with constant velocity $v$ to the right, whereas the second lattice (trapping $|\da \ra$) moves with constant velocity $-v$ to the left. This motion realizes the controlled shift. Laser pulses allow one to manipulate the internal state of the atom and to implement the Hadamard transform. For the random walk on the line the two lattices are initially tuned such that their minima overlap and the atom is placed into one of them, $x_0$. The lattices are put into motion and the Hadamard pulse is applied at times $t=nd/v$, where $d$ is the lattice spacing. A simple fluorescence measurement - together with several repetitions of the experiment - allows one to measure the resulting probability distribution. It is possible to implement a walk on the line with one or two absorbing boundaries by shining a laser at a certain position $b$ which couples both internal atomic states to a rapidly decaying level. Modifications of the trapping geometry allow to implement the walk on a circle \cite{Duer:02a}.

This set-up is simple enough that many (hundreds) steps of the random walk could be implemented with present day technology. In particular since each lattice site is subject to the same coin-transformation, the procedure does not require individual adressability of the different lattice sites, which is the big challenge when this type of implementation is used to build a quantum computer.

Hopefully in the near future either of this proposals or a new one will be implemented to begin experimental observations of  quantum walks. Furthermore it would be interesting to see an experimental proposal for the continuous time random walk in some suitable physical system.

%What about the Chinese who want to simulate the continous time walk???

\section{Decoherence} \label{Sec:decoherence}

The world of quantum mechanics shows all kinds of counter intuitive signs. Yet in the classical world of our daily experience we rarely see a cat that is in a superposition of dead and alive. The emergence of the classical world from an underlying quantum micro-world is a topic of active debate, research and speculation. Usually it is assumed that interaction with a larger environment induces decoherence, which in turn makes the world classical. We will describe how one could imagine that the classical random walk emerges form its quantum counterpart. 

The crucial difference between the quantum and the classical walk are the quantum coherences, which are present only in the first. So a transition from quantum to classical has to somehow kill these coherences. We have seen in Sec. \ref{Sec:walk} that the classical walk can be obtained from the quantum walk if we measure the coin register at every iteration. Another way to obtain the classical walk is to measure the positions at every step. We can now imagine some continuous way to interpolate between non-measuring (the quantum case) and completely measuring (the classical case) and study the transition along this pathway. An intermediate step would be to measure the register (coin or position or both) only with some probability $p_{meas}$ at each step, where $p_{meas}=0$ corresponds to the purely quantum and $p_{meas}=1$ to the purely classical case. Numerical studies of this transition from quantum to classical have been executed for the circle, line and hypercube by Kendon and Tregenna \cite{Kendon:02a} and extensive analytical expressions for the walk on the line with decoherence of the coin-space have been obtained by Brun et al. \cite{Brun:02a}\footnote{We have seen that the ion trap implementation \cite{Travaglione:02a} of quantum walks mentioned in Sec. \ref{Sec:implementation} is especially well suited to study coin-decoherence.}.

They find that the classical behavior sets in very rapidly. As an indicator we can look at the variance of the walk. In the purely quantum case we have seen that $\sigma^2 \sim T^2$, whereas in the classical case $\sigma ^2 \sim T$. For the walk on the line with coin decoherence Brun et al. \cite{Brun:02a,Brun:02c} compute the variance as a function of the transition parameter $p_{meas}$ to find that even for very small values of $p_{meas}$ (very 'close' to quantum) the variance is already linear. Kendon's numerical results for both coin and position decoherence confirm that for all $p_{meas} > 1/T$ the behavior of the variance is essentially classical (linear in $T$). One could have expected that the variance would level off gradually with increasing $p_{meas}$, but this does not seem to be the case. 

There is another interesting option to transition the coined walk between quantum and classical taken by Brun et al. \cite{Brun:02a,Brun:02b}. Imagine the quantum discrete time random walk with the modification that at each iteration we use a fresh coin Hilbert space to flip our coin. This does not allow the quantum coherences between the coin degrees of freedom to ever interfere and hence the walk behaves like the classical walk. A way to go from quantum to classical would be to use a new coin space only once in a while; a walk alternating between two coin-spaces at every step can be thought of close to the quantum walk, a walk with three coin-spaces is already less quantum and so on to the walk with an amount of coin spaces close to $T$, which is close to classical. It turns out that this way to transition from quantum to classical induces a very different behavior of the variance $\sigma ^2$: it stays quadratic in $T$ for a whole range of coin-spaces to become linear only in the limit of one new coin per time step. So the quantum character in this transition is very robust!

Kendon and Tregenna's \cite{Kendon:02a} numerical studies also show that decoherence can improve both convergence to uniform and reliability of hitting. For small $p \ll 1/T$ (where the walk is already a 'little bit classical') under certain conditions the convergence to uniform on the circle (or line-segment) can be faster than in the purely quantum case.

There is certainly a lot more to understand about decoherence in quantum random walks than we do presently - these initial studies seem to unveil just the tip of an enormous iceberg.

{\em Conclusion:} Starting from a rather simple physical effect we have tried to build up the beautiful framework of quantum random walks. We have presented several rather striking differences in the behavior of these walks as compared to their classical counterparts concerning their distribution, their mixing and hitting properties. With the advent of quantum computation it is our hope that quantum random walks will be utilized to provide new and fast algorithms that can be run on a quantum computer. We have seen that the first steps have been made in this direction, also from the implementation point of view. This area of research is still dotted with open questions and problems and we hope the reader has obtained a good general intuition and overview as to how this exciting field has evolved so far and where it might go.

{\em Acknowledgments:} The work on this article has drawn great benefits from fruitful discussions with Dorit Aharonov, Andris Ambainis, Andrew Childs, Richard Cleve, Eddy Farhi, Vivian Kendon, Neil Shenvi, Ben Tregenna, Umesh Vazirani and especially Peter Knight.
JK's effort is sponsored by the Defense
Advanced Research Projects Agency (DARPA) and Air Force Laboratory, Air
Force Materiel Command, USAF, under agreement number F30602-01-2-0524.

\bibliographystyle{unsrt}  
  
%\bibliography{masterbib} 

\begin{thebibliography}{10}

\bibitem{Nielsen:book}
M.A. Nielsen and I.L. Chuang.
\newblock {\em Quantum Computation and Quantum Information}.
\newblock Cambridge University Press, Cambridge, UK, 2000.

\bibitem{Hughes:95a}
R.J. Hughes, D.M. Alde, P.~Dyer, G.G. Luther, G.L. Morgan, and M.~Schauer.
\newblock Quantum cryptography.
\newblock {\em Contemp. Phys.}, 36(3):149--163, May-{J}une 1995.

\bibitem{Gisin:02a}
N.~Gisin, G.G. Ribordy, W.~Tittel, and H.~Zbinden.
\newblock Quantum cryptography.
\newblock {\em Rev. Mod. Phys.}, 74(1):145--195, Jan 2002.

\bibitem{Shor:94a}
P.W. Shor.
\newblock Algorithms for quantum computation: {Discrete} log and factoring.
\newblock In S.~Goldwasser, editor, {\em Proceedings of the 35th Annual
  Symposium on the Foundations of Computer Science}, pages 124--134, Los
  Alamitos, CA, 1994. IEEE Computer Society.
\newblock Final version in~\cite{Shor:97a}.

\bibitem{Aharonov:93a}
Y.~Aharonov, L.~Davidovich, and N.~Zagury.
\newblock Quantum random walks.
\newblock {\em Phys. Rev. A}, 48(2):1687--1690, 1993.

\bibitem{Feynman:65a}
{R. P.} Feynman and {A. R.} Hibbs.
\newblock {\em Quantum mechanics and path integrals}.
\newblock International series in pure and applied physics. McGraw-Hill, New
  York, 1965.

\bibitem{Meyer:96a}
D.~Meyer.
\newblock From quantum cellular automata to quantum lattice gases.
\newblock {\em J. Stat. Phys.}, 85:551--574, 1996.

\bibitem{Meyer:96b}
D.~Meyer.
\newblock On the absence of homogeneous scalar unitary cellular automata.
\newblock {\em Phys. Lett. A}, 223(5):337--340, 1996.

\bibitem{Watrous:unp}
J.~Watrous.
\newblock unpublished, 2000.

\bibitem{Watrous:01a}
J.~Watrous.
\newblock Quantum simulations of classical random walks and undirected graph
  connectivity.
\newblock {\em Journal of Computer and System Sciences}, 62(2):376--391, 2001.

\bibitem{Aharonov:01a}
D.~Aharonov, A.~Ambainis, J.~Kempe, and U.~Vazirani.
\newblock Quantum walks on graphs.
\newblock In {\em Proc. 33th STOC}, pages 50--59, New York, NY, 2001. ACM.

\bibitem{Ambainis:01b}
A.~Ambainis, E.~Bach, A.~Nayak, A.~Vishwanath, and J.~Watrous.
\newblock One-dimensional quantum walks.
\newblock In {\em Proc. 33th STOC}, pages 60--69, New York, NY, 2001. ACM.

\bibitem{Harmin:97a}
D.A. Harmin.
\newblock Coherent time evolution on a grid of {L}andau-{Z}ener anticrossings.
\newblock {\em Phys. Rev. A}, 56(1):232--251, 1997.

\bibitem{Bouwmeester:99a}
D.~Bouwmeester, I.~Marzoli, G.P. Karman, W.~Schleich, and J.P. Woerdman.
\newblock Optical galton board.
\newblock {\em Phys. Rev. A}, 61:013410--1--9, 1999.

\bibitem{Nayak:00a}
A.~Nayak and A.~Vishwanath.
\newblock Quantum walk on a line, 2000.
\newblock {DIMACS} Technical Report 2000-43 and Los Alamos preprint archive,
  \texttt{quant-ph/0010117}.

\bibitem{Bach:02a}
E.~Bach, S.~Coppersmith, M.~Goldschen, R.~Joynt, and J.~Watrous.
\newblock One-dimensional quantum walks with absorbing boundaries, 2002.
\newblock {L}anl-archive quant-ph/0207008.

\bibitem{Moore:01a}
C.~Moore and A.~Russell.
\newblock Quantum walks on the hypercube.
\newblock In J.D.P. Rolim and S.~Vadhan, editors, {\em Proc. RANDOM 2002},
  pages 164--178, Cambridge, MA, 2002. Springer.

\bibitem{Kempe:02b}
J.~Kempe.
\newblock Quantum random walks hit exponentially faster, 2002.
\newblock lanl-ar{X}iv quant-ph/0205083.

\bibitem{Mackay:02a}
T.D. Mackay, S.D. Bartlett, L.T. Stephenson, and B.C. Sanders.
\newblock Quantum walks in higher dimensions.
\newblock {\em J. Phys. A: Math. Gen.}, 35:2745, 2002.

\bibitem{Yamasaki:02a}
T.~Yamasaki, H.~Kobayashi, and H.~Imai.
\newblock An analysis of absorbing times of quantum walks.
\newblock In C.~Calude, M.J. Dinneen, and F.~Peper, editors, {\em
  Unconventional Models of Computation, Third International Conference, UMC
  2002, Kobe, Japan, October 15-19, 2002, Proceedings}, volume 2509 of {\em
  Lecture Notes in Computer Science}, pages 315--330. Springer, 2002.

\bibitem{Tamon:02a}
C.~Tamon.
\newblock Non-uniform mixing in continuous quantum walks, 2002.
\newblock {L}anl-ar{X}iv quant-ph/0209106.

\bibitem{Shenvi:02b}
N.~Shenvi, J.~Kempe, and K.B. Whaley.
\newblock A quantum random walk search algorithm.
\newblock {\em Phys. Rev. A}, 2003.
\newblock to appear.

\bibitem{Grover:96a}
L.~Grover.
\newblock A fast quantum mechanical algorithm for database search.
\newblock In {\em Proc. 28th STOC}, pages 212--219, Philadelphia, Pennsylvania,
  1996. ACM Press.

\bibitem{Farhi:98a}
E.~Farhi and S.~Gutmann.
\newblock Quantum computation and decision trees.
\newblock {\em Phys. Rev. A}, 58:915--928, 1998.

\bibitem{Childs:01a}
A.~Childs, E.~Farhi, and S.~Gutmann.
\newblock An example of the difference between quantum and classical random
  walks.
\newblock {\em Quantum Information Processing}, 1:35, 2002.
\newblock lanl-report quant-ph/0103020.

\bibitem{Aldous:notes}
D.~Aldous and J.~Fill.
\newblock Reversible markov chains and random walks on graphs.
\newblock preprint available at
  http://stat-www.berkeley.edu/users/aldous/book.html.

\bibitem{Childs:02a}
A.M. Childs, R.~Cleve, E.~Deotto, E.~Farhi, S.~Gutmann, and D.A. Spielman.
\newblock Exponential algorithmic speedup by quantum walk, 2002.
\newblock lanl-report quant-ph/0209131.

\bibitem{Preskill:private}
J.~Preskill, 2002.
\newblock private communication.

\bibitem{Shenvi:unp}
N.~Shenvi and J.~Kempe.
\newblock unpublished.

\bibitem{Papadimitriou:book}
C.~Papadimitriou.
\newblock {\em Computational Complexity}.
\newblock Addison Wesley, Reading, Massachusetts, 1994.

\bibitem{Motwani:book}
R.~Motwani and P.~Raghavan.
\newblock {\em \em Randomized Algorithms}.
\newblock Cambridge University Press, 1995.

\bibitem{Sinclair:93a}
A.~Sinclair.
\newblock {\em Algorithms for Random Generation and counting, a Markov Chain
  approach}.
\newblock Birkhauser press, 1993.

\bibitem{Dyer:91a}
M.~Dyer, A.~Frieze, and R.~Kannan.
\newblock A random polynomial-time algorithm for approximating the volume of
  convex bodies.
\newblock {\em Journal of the ACM}, 38(1):1--17, January 1991.

\bibitem{Jerrum:01a}
M.~Jerrum, A.~Sinclair, and E.~Vigoda.
\newblock A polynomial-time approximation algorithm for the permanent of a
  matrix with non-negative entries.
\newblock In {\em Proc. 33th STOC}, pages 712--721, New York, NY, 2001. ACM.

\bibitem{Schoning:99a}
U.~{Sch\"{o}ning}.
\newblock A probabilistic algorithm for {$k$-{SAT}} and constraint satisfaction
  problems.
\newblock In {\em 40th Annual Symposium on Foundations of Computer Science},
  pages 17--19. IEEE, 1999.

\bibitem{Feller:book}
W.~Feller.
\newblock {\em An Introduction to Probability Theory and its Applications}.
\newblock John Wiley, New York, 1968.

\bibitem{Lovasz:98a}
L.~Lovasz and P.~Winkler.
\newblock Mixing times.
\newblock In D.~Aldous and J.~Propp, editors, {\em Microsurveys in Discrete
  Probability}, volume~41 of {\em DIMACS Series on Disc. Math. and Theoret.
  Comp. Sci.}, pages 85--134. AMS, 1998.

\bibitem{Bennett:73a}
Ch. Bennett.
\newblock Logical reversibility of computation.
\newblock {\em IBM J. Res. Develop.}, 17:5225, 1973.

\bibitem{Aharonov:unp}
D.~Aharonov.
\newblock unpublished.

\bibitem{Aharonov:03a}
D.~Aharonov and A.~Ta-Shma.
\newblock Adiabatic quantum state generation and statistical zero knowledge,
  2003.
\newblock to appear in {\em Proc. 35th STOC} (ACM, New York, NY).

\bibitem{Bernstein:97a}
E.~Bernstein and U.~Vazirani.
\newblock Quantum complexity theory.
\newblock {\em SIAM Journal on Computing}, 26:1411, 1997.

\bibitem{Travaglione:02a}
B.C. Travaglione and G.J. Milburn.
\newblock Implementing the quantum random walk.
\newblock {\em Phys. Rev. A}, 65:032310, 2002.

\bibitem{Sanders:02a}
B.C. Sanders, S.D. Bartlett, B.~Tregenna, and P.L. Knight.
\newblock Quantum quincunx in cavity quantum electrodynamics, 2002.
\newblock {arXive} eprint quant-ph/0207028.

\bibitem{Duer:02a}
W.~D\"ur, R.~Raussendorf, V.M. Kendon, and {H.-J.} Briegel.
\newblock Quantum random walks in optical lattices.
\newblock {\em Phys. Rev. A}, 66:052319, 2002.

\bibitem{Sasura:02a}
M.~Sasura and V.~Buzek.
\newblock Cold trapped ions and quantum information processors.
\newblock {\em J. Mod. Optics}, 49:1593--1647, 2002.

\bibitem{Wineland:98a}
D.J. Wineland, C.~Monroe, W.M. Itano, D.~Leibfield, B.E. King, and D.M Meekof.
\newblock Experimental issues in coherent quantum-state manipulation of trapped
  atomic ions.
\newblock {\em J. Res. Natl. Inst. Stand. Technol.}, 103:256--328, 1998.

\bibitem{Scully:97a}
M.O. Scully and M.S. Zubairy.
\newblock {\em Quantum Optics}.
\newblock Cambridge University Press, New York, 1997.

\bibitem{Kendon:02a}
V.~Kendon and B.~Tregenna.
\newblock Decoherence is useful in quantum walks, 2002.
\newblock {L}anl-ar{X}ive quant-ph/0209005.

\bibitem{Brun:02a}
T.A. Brun, H.A. Carteret, and A.~Ambainis.
\newblock The quantum to classical transition for random walks, 2002.
\newblock {L}ANL preprint quant-ph/0208195.

\bibitem{Brun:02c}
T.A. Brun, H.A. Carteret, and A.~Ambainis.
\newblock Quantum random walks with decoherent coins, 2002.
\newblock {L}ANL preprint quant-ph/0210180.

\bibitem{Brun:02b}
T.A. Brun, H.A. Carteret, and A.~Ambainis.
\newblock Quantum walks driven by many coins, 2002.
\newblock {L}ANL preprint quant-ph/0210161.

\bibitem{Shor:97a}
P.W. Shor.
\newblock Polynomial-time algorithms for prime factorization and discrete
  logarithms on a quantum computer.
\newblock {\em SIAM J. Comp.}, 26(5):1484--1509, 1997.

\end{thebibliography}

\vskip.8cm

%\hspace{1cm} \begin{minipage}{8cm}
%{\em Julia Kempe} obtained her masters degrees in Algebra (1995) and Theoretical Physics (1996) in Paris, France and a PhD in Mathematics (2001) from the University of California, Berkeley, USA, as well as a PhD in Computer Science (2001) from the \'Ecole Nationale Superieure des T\'el\'ecommunications in Paris, France. Since 2001 she is a researcher at the CNRS at Paris-Sud University in Orsay. Her research interests include quantum computation and quantum information theory.
%\end{minipage}

\end{document}